\pdfoutput=1
\documentclass[10pt,twocolumn,letterpaper]{article}
\usepackage[T1]{fontenc}
\usepackage[utf8]{inputenc}
\usepackage[margin=0.72in,columnsep=0.25in]{geometry}
\usepackage{lmodern,microtype}
\usepackage{amsmath,amssymb,booktabs,tabularx,array}
\usepackage{graphicx}
\usepackage[table]{xcolor}
\usepackage{tikz}
\usetikzlibrary{arrows.meta,positioning}
\usepackage{enumitem}
\usepackage{caption,sectsty}
\allsectionsfont{\raggedright}
\usepackage[hidelinks]{hyperref}
\usepackage{xurl}

\usepackage{xspace}

\newcolumntype{C}{>{\centering\arraybackslash}X}

\newcommand*{\ie}{\emph{i.e.},\@\xspace}
\newcommand*{\etc}{\emph{etc.}\@\xspace}

\numberwithin{equation}{section}

\setlist{nosep,leftmargin=*}
\newcolumntype{Y}{>{\raggedright\arraybackslash}X}
\title{\vspace{-1.5em}\textbf{MultiphaseHUB for Evidence-Grounded Metadata Curation and Task-Aware Compression of Multiphase Data}}
\author{Yingyan Zeng\textsuperscript{a}, J.P. Panda\textsuperscript{b},  Avranshu Mahanty\textsuperscript{a}, Han Hu\textsuperscript{c,d}, Depeng Xu\textsuperscript{e}, and Ying Sun\textsuperscript{b,*}\\[4pt]
\small\textsuperscript{a}Department of Mechanical, Materials, and Industrial Engineering, University of Cincinnati, Cincinnati, OH\\
\small\textsuperscript{b}Department of Mechanical Engineering and Engineering Science,\\
\small University of North Carolina at Charlotte, Charlotte, NC\\
\small\textsuperscript{c}Department of Mechanical Engineering, University of Arkansas, Fayetteville, AR\\
\small\textsuperscript{d}Arkansas Research Institute for Electronics Systems, University of Arkansas, Fayetteville, AR\\
\small\textsuperscript{e}Department of Cybersecurity, University of North Carolina at Charlotte, Charlotte, NC\\}
\date{}

\begin{document}
\maketitle

\begin{abstract}
MultiphaseHUB is a community infrastructure for finding, comparing, and reusing multiphase transport data. Its expanded public catalog indexes experimental, simulation, literature-derived, and operational data and links records to source repositories, publications, physical conditions, and access terms. Six interactive literature catalogs support cross-study comparison of reported conditions and digitized results across boiling, condensation, spray cooling, and shock--droplet interaction. Researchers can use these records to formulate hypotheses about physical regimes and mechanisms and identify datasets for testing their predictions. The public MultiphaseHUB Agent supports natural-language dataset discovery, side-by-side comparison of test conditions, and charts of reported ranges and category counts. A recorded demonstration compares a subcooled flow-boiling dataset with a saturated pool-boiling dataset and shows how differences in regime and geometry affect dataset selection. We also present large language model (LLM)-assisted metadata extraction and figure digitization, and a preliminary evaluation of modality-specific residual vector quantization on temperature, hydrophone, and video measurements from one saturated pool-boiling run on copper foam. Cross-facility harmonization, dataset valuation, and the evaluation of task-aware compression and missing-modality reconstruction remain research directions.
\end{abstract}

\section{Introduction}
\label{sec:introduction}

Multiphase transport is central to boiling and condensation, immersion cooling, drop impact, particulate deposition, and many engineering processes that couple momentum, heat, mass, and interfacial dynamics. Investigating these processes requires observations across multiple spatial and temporal scales. High-speed imaging captures interface motion, infrared thermography measures spatial thermal response, acoustic sensing detects transient events, and pressure or electrical measurements characterize operating conditions and system behavior. Experiments and simulations therefore generate diverse, multimodal, high-dimensional datasets that can support physical understanding, model development, and engineering design. Their scientific value depends on the context needed to interpret the observations, including the geometry, working fluid, surface state, boundary conditions, sensor configuration, calibration, and uncertainty.

Despite the growing availability of multiphase transport data, reuse across studies remains difficult. Raw measurements, annotations, preprocessing scripts, and experimental descriptions are often distributed across laboratory websites, publication supplements, and general-purpose repositories. Important details may appear only in a paper's methods section or in a separate calibration document. As a result, users must reconstruct the connections among data, physical conditions, and analysis procedures before determining whether a dataset is appropriate for a new question or can be used for a downstream analysis.
Differences in variable definitions, units, timestamps, sensing configurations, and operating regimes further complicate comparison. Persistent access to the measurement data is an essential starting point, but physically meaningful reuse also requires sufficient information about what was measured, how it was measured, and under which conditions.

The volume and incompleteness of multiphase transport measurements create coupled barriers to reuse. On one side, high-speed videos, temperature measurements, and acoustic signals can be costly to transfer and analyze, particularly when a user needs only a subset of their information. On the other side, these measurements may also contain missing segments or lack entire modalities because of sensor limitations, acquisition failures \etc

We introduce MultiphaseHUB~\cite{hub} as a community infrastructure and an integrated framework for organizing, discovering, and reusing multiphase transport data. The infrastructure supports access to distributed data resources and their associated metadata, publications, analysis tools, and derived features, while retaining source repository links, provenance, and reuse boundaries. The framework integrates literature-grounded metadata curation and scientific data compression to support the interpretation and efficient reuse of these resources across physical systems and sensing modalities. Its development follows the goals of findability, accessibility, interoperability, and reusability articulated by the FAIR principles~\cite{fair}, with particular attention to the physical context and measurement information required for multiphase analysis.

The framework brings together dataset discovery and comparison, literature analysis and data curation, and scientific data compression. The expanded catalog supports faceted search across resources with different operating conditions, measured variables, and access terms. The public MultiphaseHUB Agent searches the catalog in response to natural-language queries, compares test conditions, and visualizes reported parameter ranges and category counts. The literature-analysis component provides a complementary way to compare findings reported in papers. It extracts metadata from text and tables and digitizes quantitative results from figures, linking extracted values to source evidence for verification. By comparing trends across operating conditions, researchers can identify candidate regime boundaries, formulate hypotheses about the processes that govern each regime, and find measurements with which to test their predictions.
Meanwhile,  a modality-specific scientific compression method is proposed as one component in MultiphaseHUB with cross-modal imputation, using available measurements and physical context to estimate missing signals. The resulting proxies (\ie compressed representations) would provide compact representations of observed measurements together with estimates of missing information. Their scientific usefulness depends on both reconstruction fidelity and imputation reliability. 
The extracted metadata from the agentic literature analyzer connects these objectives by describing operating conditions and measurement availability, helping determine which quantities matter and when relationships across modalities can support imputation.
Together, these components support a workflow from identifying relevant data and interpreting their experimental conditions to constructing compact representations for subsequent analysis. They also provide a basis for investigating how metadata reliability and representation error affect downstream scientific utility.

We describe the expanded public catalog and illustrate how the MultiphaseHUB Agent compares experimental conditions across datasets. We then describe the literature extraction methods and present preliminary scientific compression results from a single saturated pool-boiling run. The research roadmap includes evaluation across facilities, task-aware retrieval and dataset valuation, and conditional reconstruction of missing measurements from available modalities and physical context.

\section{Related Data and AI Efforts}
\label{sec:related}

Open multiphase data resources already span experimental images, annotations, simulations, and machine-learning benchmarks. BubbleID couples boiling imagery, annotations, trained models, and tutorials for interface-dynamics analysis~\cite{bubbleid,bubbleiddata}. Hessenkemper \emph{et al.} released annotated and synthetic bubble images with a persistent dataset identifier to support image-based bubble identification~\cite{hessenkemper}. BubbleML provides simulated boiling fields, baseline scientific-machine-learning tasks, and reproducibility materials for multiphysics learning~\cite{bubbleml}. While these efforts demonstrate that useful multiphase resources can be built around a well-defined physical system and task, they also expose a persistent interoperability problem: data scope, variables, formats, labels, sensor conditions, and quality controls differ substantially across resources.

\begin{figure*}[t]
  \centering
  \includegraphics[width=0.78\textwidth]{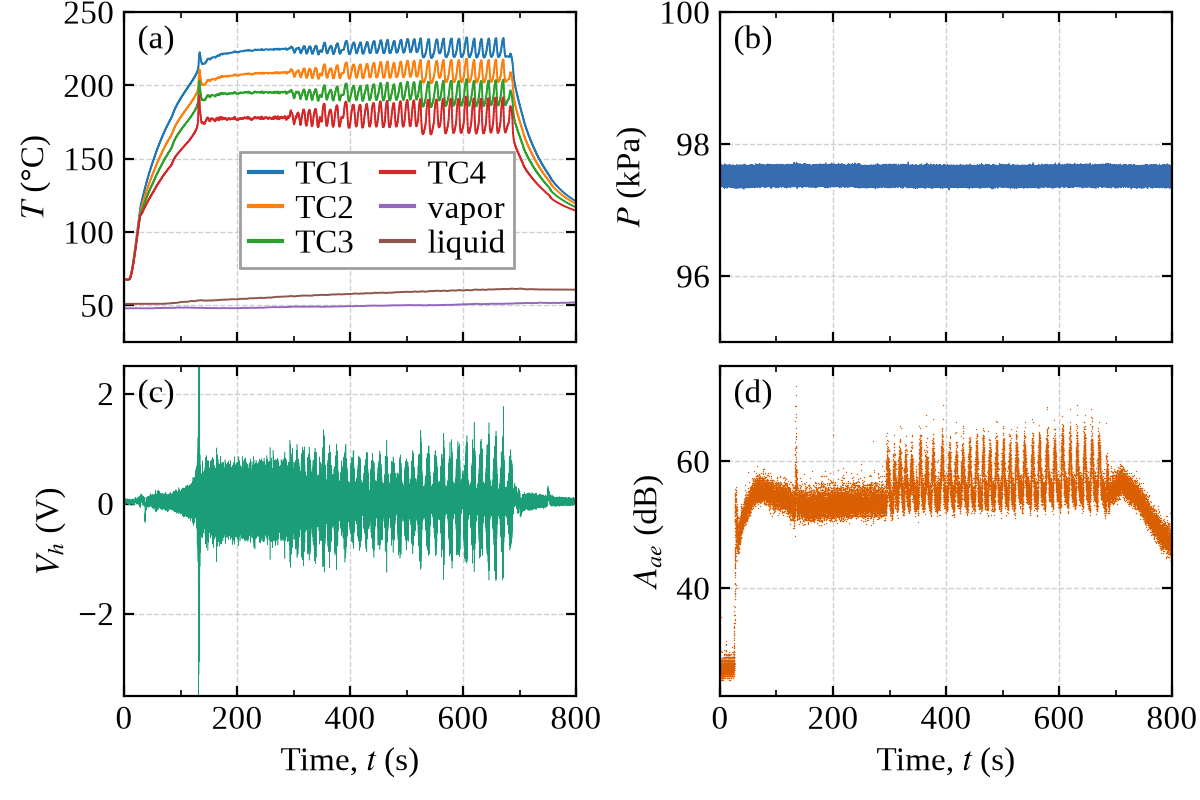}\par\smallskip
  \begin{minipage}{0.88\textwidth}
    \raggedright
    {\small (e)}\par
    \includegraphics[width=\linewidth]{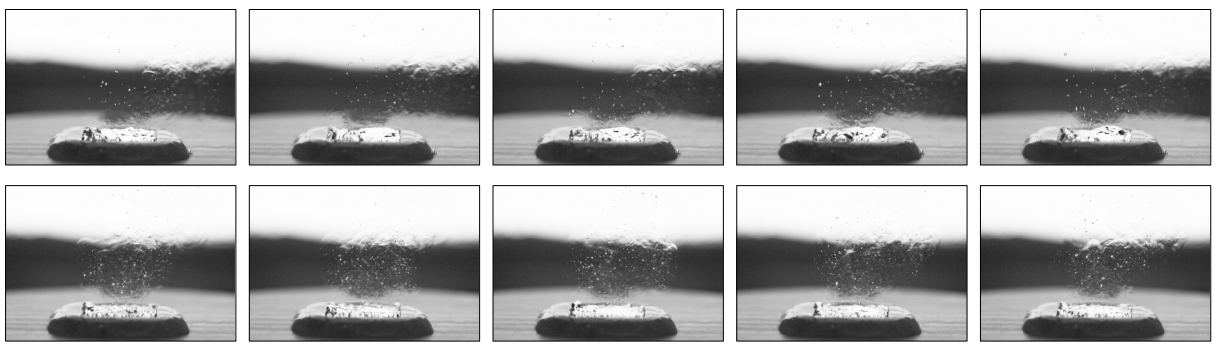}
  \end{minipage}
  \caption{Raw measurements and high-speed video frames from the subcooled flat-copper pool-boiling run in BoilingBench-1. (a) Copper-block and fluid temperatures. (b) Pressure. (c) Hydrophone voltage. (d) Acoustic-emission hit amplitude. (e) Selected frames from the high-speed video.}
  \label{fig:bb-raw}
\end{figure*}

\begin{figure}[t]
  \centering
  \includegraphics[width=\columnwidth]{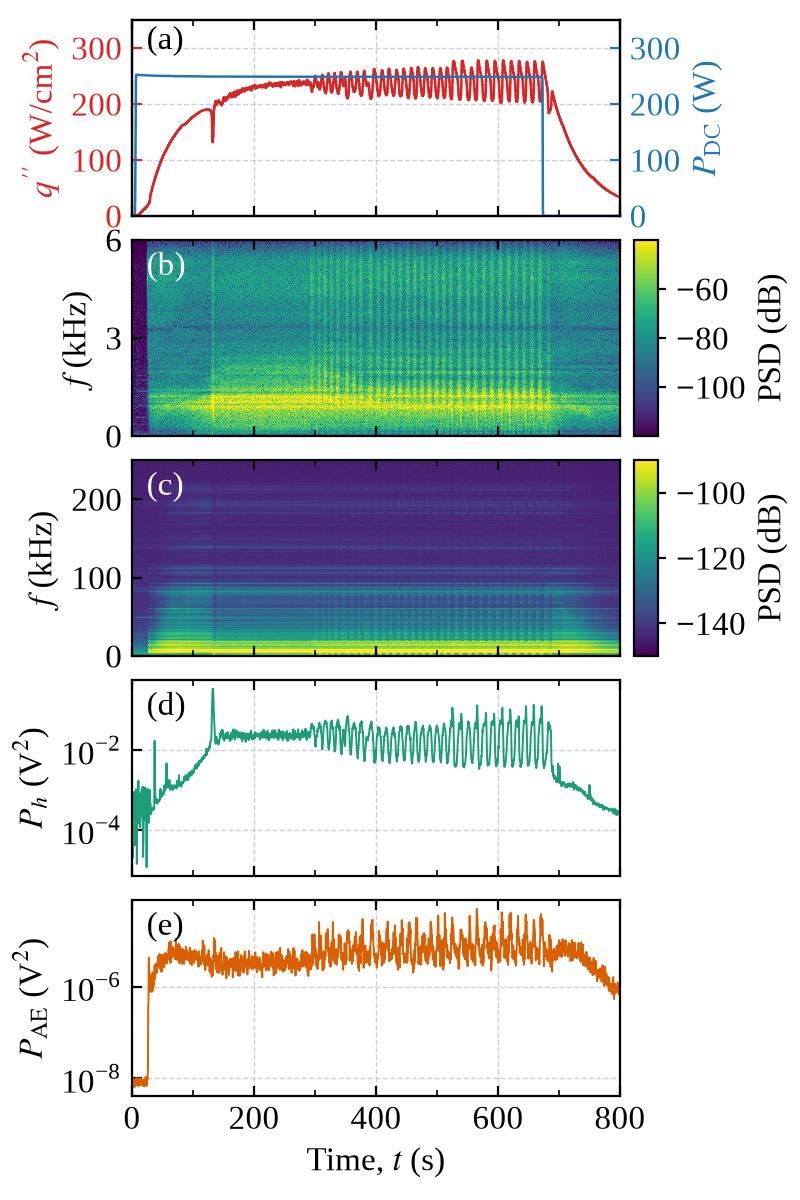}
  \caption{Processed results from the BoilingBench-1 run shown in Fig.~\ref{fig:bb-raw}. (a) Heat flux estimated from the temperature gradient and measured DC input power. (b) Hydrophone spectrogram. (c) Acoustic-emission spectrogram. (d) Band-integrated hydrophone power. (e) Band-integrated acoustic-emission power. PSD denotes power spectral density.}
  \label{fig:bb-processed}
\end{figure}

BoilingBench-1~\cite{hu2026boilingbench} illustrates the distinction between source measurements and derived data. It is an ambient-pressure, subcooled pool-boiling run on a flat copper surface. The records are aligned to the thermocouple clock. Fig.~\ref{fig:bb-raw}a--d shows the temperature, pressure, hydrophone voltage, and acoustic-emission (AE) hit amplitude, and Fig.~\ref{fig:bb-raw}e shows frames from the high-speed video. Fig.~\ref{fig:bb-processed}a plots the derived heat flux alongside the measured DC input power. The heat flux is estimated from a linear fit to the thermocouple readings. Fig.~\ref{fig:bb-processed}b--e shows the hydrophone and AE spectrograms and the corresponding band-integrated powers.

The heater power is held approximately constant from 5~s until it is switched off at about 670~s (Fig.~\ref{fig:bb-processed}a). Oscillatory boiling and acoustic activity persist briefly after shutoff as sensible heat stored in the copper block continues to transfer to the water.

In recent years, AI methods are increasingly applied to segmentation, tracking, optical-flow estimation, surrogate modeling, and nonintrusive sensing in multiphase systems. For example, BubbleID supports image-based analysis of bubble interface dynamics~\cite{bubbleid}, BubbleML provides simulation datasets and benchmarks for scientific machine learning~\cite{bubbleml}, and Wallen \emph{et al.} combine acoustic-emission sensing with unsupervised learning to investigate flow condensation~\cite{wallen}. These efforts illustrate how different data modalities and analysis methods can support the study of multiphase phenomena. However, applying an existing method to a new dataset requires understanding the physical conditions, sensing configuration, preprocessing, and evaluation procedures used in its development. Differences in these factors can affect whether the method remains suitable and how its reported performance should be interpreted. This motivates connecting datasets with associated AI tools, publications, and experimental context within MultiphaseHUB, helping users identify relevant analysis methods and determine what adaptation or additional validation may be needed for their intended use.

The development of datahubs and data banks has been demonstrated\ in turbulence, materials science, structural biology, genomics, and other research fields. The Johns Hopkins Turbulence Databases provide programmatic access to large simulation fields through virtual-sensor services rather than requiring users to handle the entire archive~\cite{jhtdb}. The Materials Project combines structured material-property records with programmatic access~\cite{materialsproject}. The Protein Data Bank demonstrates community deposition, curation, standardized representation, and rich search for experimentally determined structures~\cite{pdb}. The Genomic Data Commons harmonizes cancer genomic and associated clinical data and supports cross-study analysis under explicit access controls~\cite{gdc}. Dataverse provides persistent identification, citation, versioning, and collection-level organization for research data~\cite{dataverse}.
Together, these dataset and datahub efforts demonstrate the value of combining reliable data access with scientific context and analysis tools. They motivate MultiphaseHUB as a community infrastructure and integrated framework that builds on existing repositories to support the discovery, interpretation, and reuse of multiphase data across physical systems and sensing modalities. The FAIR principles provide a foundation for this effort by emphasizing persistent identifiers, descriptive metadata, provenance, and clear access and reuse information~\cite{fair}.

Literature-grounded AI offers a way to recover relevant context when experimental conditions and measurement definitions are distributed across papers, supplements, and repository descriptions. Retrieval-augmented generation (RAG) combines document retrieval with language generation, allowing an output to be conditioned on external evidence~\cite{rag}.
Agentic approaches extend this capability by selecting subsequent actions in response to intermediate observations.
ReAct, for example, interleaves reasoning with actions that access external information~\cite{react}. For multiphase metadata curation, these capabilities motivate workflows that retrieve evidence for a requested field, extract a candidate value, check units and experimental scope, and revisit the sources when information is missing or conflicting. However, retrieving a relevant passage does not establish that a value belongs to the correct run, surface, or operating condition. The domain-specific challenge is to produce structured assertions with traceable source locations, explicit unresolved states, and reviewable validation outcomes. Accordingly, a metadata extraction tool should be evaluated through field-level correctness, evidence support, and curation effort.

Scientific data compression techniques have been developed to reduce the storage and transfer costs of high-dimensional experimental measurements and simulation outputs, facilitating data access and reuse.
Compression methods such as the fixed-rate floating-point representation introduced by Lindstrom reduce the size of numerical arrays while supporting efficient access~\cite{zfp}. Learned approaches jointly optimize an encoder, quantized representation, and decoder through rate-distortion objectives, as demonstrated by Ball\'{e} \emph{et al.} for image compression~\cite{balle}. Scientific utility nevertheless depends on the analysis performed on the reconstructed data. Laney \emph{et al.} explicitly examined the effects of lossy compression using physically motivated simulation metrics~\cite{laney}.
These studies highlight the importance of evaluating scientific compression through both reconstruction error and preservation of physical quantities of interest (QoIs), i.e., scientific quantities computed from the data for a downstream analysis, such as the projected bubble area extracted from high-speed boiling video. However, preserving these quantities in multiphase transport remains challenging because bubble-interface geometry, event timing, acoustic spectral features, and thermal statistics can exhibit different sensitivities to compression across modalities and operating conditions. Missing signal segments or unavailable modalities further limit the information needed to estimate these quantities reliably. Motivated by these challenges, we propose to couple modality-specific compression with cross-modal imputation in MultiphaseHUB, using available measurements and physical context to construct compact proxies containing reconstructed observations and estimates of missing signals.

Together, these efforts establish MultiphaseHUB~\cite{hub} as a community infrastructure and integrated framework for connecting distributed multiphase resources with literature-grounded physical metadata and compact representations whose scientific utility can be assessed. This integration provides a basis for investigating whether a dataset is suitable for an intended analysis, whether compression preserves the physical information that analysis requires, and how uncertainty in estimates of missing signals affects scientific conclusions. These questions connect the present curation and compression components to a broader research roadmap for reliable reuse and cross-study discovery.

\section{MultiphaseHUB Architecture and Core Components}
\label{sec:architecture}
\label{sec:records}

\subsection{Functional components and information flow}
MultiphaseHUB organizes three functional components (Table~\ref{tab:components}). The public dataset catalog provides discovery and source metadata. The MultiphaseHUB Agent queries catalog records, compares documented conditions, and summarizes coverage. The literature component curates evidence from publications and presents it in searchable study collections. The compression component produces compact representations of high-dimensional measurements and has been evaluated preliminarily on one boiling run.

\begin{table*}[t]
\centering\small
\caption{MultiphaseHUB components and the status of their outputs.}
\label{tab:components}
\begin{tabularx}{\textwidth}{@{}>{\raggedright\arraybackslash}p{0.19\textwidth}YY@{}}
\toprule
\textbf{Component} & \textbf{Modules or functions} & \textbf{Outputs and scope in this paper}\\
\midrule
Dataset discovery and comparison & Faceted catalog and public Agent for natural-language queries. & Source-linked records, side-by-side condition comparisons, and charts of reported ranges and counts (Sec.~\ref{subsec:dataset-comparison}).\\
Literature analysis and data curation & LLM-assisted metadata extraction, figure digitization, unit normalization, and evidence checks. & Public topic catalogs and a boiling-curve compilation.\\
Scientific data compression & Modality-specific encoding, residual vector quantization, and reconstruction. & Preliminary rate and reconstruction results on one saturated pool-boiling run.\\
\bottomrule
\end{tabularx}
\end{table*}

\subsection{Physical context, metadata, and provenance}
Table~\ref{tab:schema} summarizes the information supporting these components, including source identity, physical conditions, measurement semantics, derived artifacts, and provenance. These categories describe the information needed for interpretation and reuse.

\begin{table*}[t]
\centering\small
\caption{Metadata categories supporting physical interpretation and reuse of multiphase data. Representative contents describe the information needed by the components; availability and completeness vary across resources.}
\label{tab:schema}
\begin{tabularx}{\textwidth}{@{}>{\raggedright\arraybackslash}p{0.18\textwidth}YY@{}}
\toprule
\textbf{Field group} & \textbf{Representative contents} & \textbf{Role in reuse}\\
\midrule
Identity and source & Dataset identifier, source URL, study/case ID, asset version, citation & Resolve the original resource and distinguish related releases.\\
Physical conditions & Fluid, geometry, surface state, pressure, heat flux, mass flux, boundary conditions & Establish the operating regime and limits of comparison.\\
Measurement semantics & Variable definitions, units, modality, sampling, calibration, synchronization, uncertainty & Interpret each observation and identify incompatible measurements.\\
Derived artifacts & Compiled numerical series, features or QoIs, processing code, codec version, reconstruction settings, evaluation scope & Reproduce transformations and assess utility of derived data.\\
Evidence and access & Source page/table/figure, extraction method, field status, reviewer, license, access conditions & Trace assertions and preserve the source's reuse conditions.\\
\bottomrule
\end{tabularx}
\end{table*}

Each metadata value should be linked to the study, experimental case, measurement, or time interval it describes. Derived features and compression results should also identify their source data and processing settings. These links support data selection, reproducible analysis, and comparison across studies.

\subsection{Public catalog and access}
\label{subsec:catalog}
The public catalog~\cite{hub} supports search and filtering by phenomenon, fluid, data type, and physical conditions. It indexes experiments, simulations, literature-derived compilations, and operational telemetry. On November 25, 2026, the homepage reported 156 datasets across 13 folders, including 13 hosted datasets~\cite{hub}. The catalog links to resources in Zenodo, Mendeley Data, NASA repositories, and other sources. Source repositories govern access to the original files. Open, restricted, and forthcoming resources have different access conditions. Records can link to source DOIs, papers, code, data dictionaries, and license information where available.

Table~\ref{tab:catalog-examples} illustrates differences in evidence and reuse scope across records. BubbleID, for example, pairs imagery and annotations with study documentation and a Dryad identifier~\cite{bubbleid,bubbleiddata}. Literature-derived curves, simulation fields, and measured signals have different provenance even when they describe the same physical phenomenon. Each derived artifact should identify its input records and the transformation that produced it.

\begin{table*}[t]
\centering\small
\caption{Representative records and access paths in the expanded catalog. A source DOI resolves the original archive or a named part of it.}
\label{tab:catalog-examples}
\begin{tabularx}{\textwidth}{@{}>{\raggedright\arraybackslash}p{0.23\textwidth}YY@{}}
\toprule
\textbf{Resource} & \textbf{Evidence and scope} & \textbf{Source and access}\\
\midrule
BoilingBench-Multimodal Lite & Multimodal measurements, boiling imagery and derived products across several benchmark tracks. & Open Lite archive on Zenodo~\cite{hu2026boilingbench}.\\
Pool boiling: acoustic, optical and thermal records & Saturated pool-boiling experiments across five surfaces and two thermal loading regimes. & \href{https://doi.org/10.7910/DVN/6GLGC6}{Harvard Dataverse}. Study described by Pandey et al.~\cite{pandey2024multimodal}.\\
BubbleML & Simulated boiling fields for scientific machine learning and benchmark tasks. & Open simulation data and benchmark code through the \href{https://github.com/HPCForge/BubbleML}{project repository}~\cite{bubbleml}.\\
Two-phase direct-to-chip server cooling & Controlled server wind-tunnel trials with temperatures, power, airflow, and refrigerant-side measurements. & Open \href{https://doi.org/10.5281/zenodo.23237930}{Zenodo archive} with case table and data dictionary.\\
Marconi100 operational telemetry & Measured system and node telemetry, with measurements at multiple physical scales. & \href{https://zenodo.org/records/7588815}{First Zenodo deposit} in a multi-part archive linked by the catalog.\\
\bottomrule
\end{tabularx}
\end{table*}

\subsubsection{Dataset roles in scientific workflows}
\label{subsec:dataset-roles}

A dataset's role depends on what a particular version or subset contributes to a scientific question. The same dataset can suggest a hypothesis in one study and provide a reference for another. Comparing records across studies requires checking that the quantity definitions, units, acquisition methods, and operating conditions are appropriate for the question. MDS-Onto provides a precedent for describing the experimental context in interoperable metadata, and KROMA illustrates how concepts can be matched across vocabularies~\cite{rajamohan2025mdsonto,nguyen2025kroma}.

\paragraph{Evidence for generating hypotheses.}
Measurements and results extracted from publications can reveal recurring trends, departures from them, or conditions under which the observed behavior changes. For example, a change in the slope of reported pool-boiling curves can suggest a transition worth investigating. The source, measurement or extraction method, operating conditions, and available uncertainty help determine which explanations the observations can support.

\paragraph{Reference data and hypothesis tests.}
Reference data provide a documented baseline or target, such as a boiling curve for a specified fluid, pressure, and surface. Independent measurements can test a prediction under conditions chosen for the question. For example, a hypothesis that surface condition shifts a boiling transition could be tested with complete runs using the same fluid, pressure, geometry, and heating protocol. Synchronized images and thermal measurements could show whether the thermal transition is accompanied by the predicted change in bubble behavior.

\paragraph{Model development and data provenance.}
When an analysis uses a predictive model, training data fit it, validation data guide its selection or tuning, and test data assess its predictions on held-out cases. Recording the dataset version and the split unit, such as an entire experimental run, supports reproducible evaluation~\cite{akhtar2024croissant,kapoor2023leakage}. Synthetic data can help examine how changing one factor affects a result under the assumptions of the generating model. Their provenance identifies the generating model and inputs. Access terms govern availability.

Appendix~\ref{app:dataset-roles} summarizes the inventory's intended uses, including model development and evaluation.

\subsection{Relationships among catalog, literature, and compression workflows}
Figure~\ref{fig:workflow} connects the catalog, literature, and compression workflows. Dataset records link to source repositories and to publications that document the data, where available. Literature extraction supplies evidence and quantitative results for comparison across publications. The compression study uses multimodal measurements from dataset repositories and produces compact representations with reconstruction and rate information. Source metadata can help define future tests of scientific utility, for example by specifying the conditions used to select an evaluation dataset.

\begin{figure*}[t]
\centering
\begin{tikzpicture}[
    card/.style={anchor=west, draw=black!35, rounded corners=3pt,
                 line width=0.45pt, align=left, font=\small\fontfamily{ptm}\selectfont,
                 inner xsep=6pt, inner ysep=4pt, minimum height=1.42cm},
    input/.style={card, text width=3.9cm, fill=black!3},
    method/.style={card, text width=4.6cm, fill=#1!6, draw=#1!55},
    result/.style={card, text width=5cm, fill=#1!6, draw=#1!55},
    title/.style={anchor=south west, inner sep=0pt, font=\small\fontfamily{ptm}\selectfont\bfseries},
    flow/.style={-{Latex[length=2mm]}, draw=black!70, line width=0.65pt,
                 rounded corners=2pt},
    provenance/.style={flow, densely dashed},
    link label/.style={font=\footnotesize\fontfamily{ptm}\selectfont, text=black!75, inner sep=1pt}
]
\colorlet{workflowDiscovery}{blue!50!black}
\colorlet{workflowLiterature}{teal!70!black}
\colorlet{workflowCompression}{orange!55!black}

\node[title, text=workflowDiscovery] at (6.15,0.94)
    {Dataset discovery and comparison (Sec.~\ref{subsec:dataset-comparison})};
\node[input] (repositories) at (0.55,0)
    {\textbf{Dataset repositories}\\Dataset documentation\\and data files};
\node[method=workflowDiscovery] (catalog) at (6.15,0)
    {\textbf{Public dataset catalog}\\Source-linked records\\Search and filters};
\node[result=workflowDiscovery] (agent) at (12.17,0)
    {\textbf{MultiphaseHUB Agent}\\Condition comparisons\\Range and count charts};

\node[title, text=workflowLiterature] at (6.15,-1.32)
    {Literature analysis and data curation (Sec.~\ref{sec:agent})};
\node[input] (publications) at (0.55,-2.26)
    {\textbf{Publications}\\Text, tables, and figures};
\node[method=workflowLiterature] (extraction) at (6.15,-2.26)
    {\textbf{LLM-assisted extraction}\\Metadata from text and tables\\Figure digitization};
\node[result=workflowLiterature] (literature) at (12.17,-2.26)
    {\textbf{Public AgenticTool catalogs}\\Searchable study records\\Study benchmarking};

\node[title, text=workflowCompression] at (6.15,-3.58)
    {Scientific data compression (Sec.~\ref{sec:compression})};
\node[input] (measurements) at (0.55,-4.52)
    {\textbf{Multimodal measurements}\\Temperature, video,\\ hydrophone audio, \etc};
\node[method=workflowCompression] (compression) at (6.15,-4.52)
    {\textbf{Modality-specific RVQ-VAEs}\\Compact representations\\(RVQ indices)};
\node[result=workflowCompression] (evaluation) at (12.17,-4.52)
    {\textbf{Preliminary compression results}\\Reconstruction examples\\and compression ratios};

\draw[flow] (repositories.east) -- (catalog.west);
\draw[flow] (catalog.east) -- (agent.west);
\draw[flow] (publications.east) -- (extraction.west);
\draw[flow] (extraction.east) -- (literature.west);
\draw[flow] (repositories.west) -- (0.02,0) |- (measurements.west);
\draw[flow] (measurements.east) -- (compression.west);
\draw[flow] (compression.east) -- (evaluation.west);

\draw[provenance] ([xshift=-12pt]publications.north east)
    -- ++(0,0.59) -| ([xshift=10pt]catalog.south west);
\node[link label, anchor=east] at (4.33,-1.28) {Publication link};
\end{tikzpicture}
\caption{Information flow in MultiphaseHUB. Repositories provide both source records for the public dataset catalog and measurement files for scientific compression. The dashed arrow marks the association between a publication and its dataset record when available. AgenticTool catalogs present study records and digitized series for cross-study comparison. The compression results come from one saturated pool-boiling run. Section~\ref{sec:roadmap} describes planned evaluation of how compression affects scientific quantities of interest.}
\label{fig:workflow}
\end{figure*}

\section{Literature Analysis and Data Curation}
\label{sec:agent}
This component combines two complementary modules: agentic extraction of metadata from document text and tables, and an agentic literature analyzer with an agentic digitizer for quantitative values from figures. The former describes the physical system and reported conditions; the latter recovers numerical evidence for comparison around a shared research question. Both modules draw on the same underlying language model, so the choice of backend that serves the extraction stage is made once, upstream of either module, and applies uniformly across them.

\subsection{Metadata extraction from text and tables}

The prototype extracts paper-level metadata from literature
associated with multiphase heat-transfer datasets. Each PDF is
converted into normalized text with a page and section index.
Four fixed, schema-constrained language-model passes identify
the paper type, extract numerical operating envelopes, assign
taxonomy facets, and identify stated or inferred applications:
\begin{equation}
\widehat{\mathcal M}_{d}^{(j)}
=
F_{\theta}(T_d,\mathcal V,P_j;\mathcal S_j),
\qquad j=1,\ldots,4,
\end{equation}
where $\widehat{\mathcal M}_{d}^{(j)}$ is the structured metadata
record returned by pass $j$ for document $d$, $F_{\theta}$ is the
language model with fixed pretrained parameters $\theta$,
$T_d$ is the extracted document text, $\mathcal V$ is the
controlled vocabulary, $P_j$ is the stage-specific instruction,
and $\mathcal S_j$ is the output schema. Each pass receives the
full extracted text.

For example, the record for Allred et al.~\cite{Allred2018Boiling}
identifies water as the working fluid, pool boiling as the
phenomenon, and experimental investigation as the method.
Surface facets capture complementary attributes, including
nanowire structures and superhydrophobic wettability.
Table~II of \cite{Allred2018Boiling} reports static Cassie-state contact
angles of $158^\circ$ and $165^\circ$ for the microstructured
and nanostructured surfaces, respectively. These values can
populate a paper-level contact-angle summary, while their
association with different surfaces remains essential for
subsequent interpretation.


The metadata extraction corpus now spans 154 multiphase heat-transfer papers across six phenomenon-specific catalogs. Table~\ref{tab:multiphase_summary} summarizes taxonomy assignments in that corpus. In terms of underlying transport phenomena, flow boiling and pool boiling account for the largest shares of analyzed papers, with substantial additional coverage dedicated to evaporation (\ie spray cooling), flow condensation, and shock--droplet interaction.
The working fluids evaluated across these studies reflect prominent industrial heat-transfer media, led by water and dielectric liquids, alongside various hydrofluorocarbon (HFC), hydrofluoroolefin (HFO), and hydrochlorofluorocarbon (HCFC) refrigerants, carrier gases, cryogenic fluids, organic liquids, and engineered nanofluids. From an application perspective, electronics thermal management constitutes the largest domain, followed by aerospace cooling architectures, fundamental thermal-fluid investigations, nuclear thermal-hydraulics, propulsion and combustion, heating, ventilation, air conditioning, and refrigeration (HVAC\&R), high-speed flight, power generation systems, cryogenics, and chemical process engineering. Methodologically, the cataloged investigations are overwhelmingly experimental, commonly recording heat transfer coefficients ($h$), wall superheat ($\Delta T_{\text{sat}}$), complete boiling curves, and critical heat flux (CHF), complemented by a smaller set of numerical simulations, correlation-only analyses, and review compilations.

In addition to primary physical  classifications of papers, the agentic literature-analyzer derives standardized surface classifications, thermal performance metrics, and derived tags to facilitate cross-dataset comparative analysis. Surface geometries within the catalog span plain surfaces, roughened surfaces, nanostructured surfaces, porous coatings, structured channels, and wettability-engineered surfaces. Key measured quantities center on thermal response indicators, enabling direct evaluation of boiling transport efficiency and critical operational limits across diverse experimental configurations. For example, measured variables include local heat transfer coefficients ($h$) across microchannel test sections, wall superheat ($\Delta T_{\text{sat}} = T_{\text{wall}} - T_{\text{sat}}$) values across boiling regimes, complete boiling curves ($q''$ vs. $\Delta T_{\text{sat}}$), and CHF thresholds marking the departure from nucleate boiling. Furthermore, derived tags systematically classify parameter regimes, surface modifications, and fluid attributes, streamlining data discovery, feature extraction, and machine learning pipeline integration. Examples of these derived classifications include operational regime tags (e.g., subcooled vs. saturated flow, buoyancy-driven vs. forced convection), surface morphology identifiers (e.g., sintered powder porous coatings, micro-grooved channels, nanostructured micropillars), and cross-modal acoustic-thermal features (e.g., acoustic emission power spectrum principal components and dominant bubble detachment frequencies).

The model is instructed to report numerical values in their
original units, leave unsupported fields empty, and attach
source quotations to evidence-bearing entries. A deterministic
locator maps quotations to text spans, pages, and sections
using normalized exact matching and a thresholded fuzzy
fallback. Unmatched quotations are marked unresolved.
Locating a quotation establishes traceability; assessing
whether it supports the assigned field requires additional
validation.


\begin{table*}[t]
\centering
\caption{Summary of MultiphaseHub Dataset Attributes Across Primary Categories}
\label{tab:multiphase_summary}
\small
\begin{tabularx}{\textwidth}{@{} X X X X @{}}
\toprule
\textbf{Phenomenon} & \textbf{Working Fluid} & \textbf{Application} & \textbf{Method} \\
\midrule
Flow boiling (46) & Water (70) & Electronics thermal (87) & Experimental (132) \\
Pool boiling (37) & Dielectric (60) & Aerospace (36) & Model / Simulation (6) \\
Evaporation (40) & Refrigerant HFC (17) & Fundamental (16) & Numerical / CFD (5) \\
Flow condensation (20) & Carrier gas (11) & Nuclear (11) & Correlation only (3) \\
Shock--droplet interaction (11) & Cryogen (9) & Propulsion / combustion (10) & Review compilation (1) \\
 & Organic liquid (7) & HVAC\&R (7) & \\
 & Refrigerant HFO (3) & High-speed flight (7) & \\
 & Natural refrigerant (3) & Power generation (4) & \\
 & Refrigerant HCFC (2) & Cryogenics (2) & \\
 & Nanofluid (2) & Process (2) & \\
 & Mixture (1) & Electric machine thermal (1) & \\
 & Liquid metal (1) & Materials processing (1) & \\
\bottomrule
\end{tabularx}
\par\smallskip
\begin{minipage}{\textwidth}
\footnotesize
Note: Counts are summed across the catalog's six phenomenon-specific sub-catalogs (154 papers total: 26 pool boiling, 38 flow boiling, 20 condensation, 19 capillary boiling, 40 spray cooling, 11 shock--droplet interaction). Several datasets carry more than one working-fluid, application, or method tag, so these columns' counts exceed 154. The condensation sub-catalog has no application facet and leaves about seven records without an explicit top-level method label, so the Method column total is correspondingly short of 154.
\end{minipage}
\end{table*}

\subsection{Agentic literature-analyzer: quantitative extraction from figures}
\label{subsec:literature_compiler}

The agentic literature-analyzer with agentic digitizer have been developed to capture quantitative values from published figures and express them in canonical units. Its output complements paper-level metadata by providing numerical series, such as paired wall-superheat and heat-flux values, that can be examined across publications addressing the same research problem. Recovering such a series requires identifying the plotted variables and the relevant curve or markers, interpreting the axes, and retaining the association between the extracted values and their source.

For scientific reuse, each compiled series should retain the original variable names and units, converted values and canonical units, source paper and figure or panel, series label, and extraction method. Physical metadata should identify the relevant specimen and operating conditions where reported. Extraction confidence or review status should remain distinguishable from the measurement uncertainty reported by the source. Expressing curves in common units facilitates comparison, while differences in geometry, surface state, pressure, and measurement definitions still require explicit consideration.

The compiler can serve as a callable module within a future agentic workflow: document analysis identifies the quantitative evidence needed, figure extraction recovers the corresponding values, and validation checks their units, experimental scope, and source association. The present figure-based application below illustrates the compiler's role without assuming that this orchestration has already been implemented end to end.

\subsection{Validation and human review}

Schema validation checks output structure and types.
Deterministic postprocessing converts recognized units to SI,
screens numerical values against predefined plausibility
bounds, and computes derived quantities and tags. Values
failing these bounds are quarantined from the corresponding
calculations. An optional language-model audit records
support or contradiction for populated fields.

For example, Chu et al.~\cite{Chu2013CuO} report a CHF of
approximately $250\,\mathrm{W\,cm^{-2}}$ for their
best-performing hierarchical surface. The corresponding
SI value is $2.5\times10^{6}\,\mathrm{W\,m^{-2}}$.
CHF is represented as a measured quantity, while pool
boiling remains the phenomenon. The same record illustrates
vocabulary extension: the extractor proposes
\texttt{copper\_oxide\_nanostructures} under the existing
nanostructured-surface category, and the proposal enters
the review queue.

The current queue prioritizes proposed vocabulary, selected
facet assignments without located evidence, inferred or
lower-confidence applications, and quarantined values.
Flagged metadata remain accessible. Automated confirmation
flags indicate triage outcomes rather than completed expert
review, and model-reported confidence is not interpreted
as a calibrated probability of correctness.

For figure-derived values, review must also check the axis scale, curve or marker assignment, variable definitions, and conversion from plotted units to canonical units. Digitization error should be evaluated separately from the uncertainty of the original measurements.

Human review addresses ambiguities that numerical checks
cannot resolve. In Allred et al.~\cite{Allred2019Petal},
Table~1 reports a static contact angle of $144^\circ$,
a receding angle of approximately $30^\circ$, and an
advancing angle of $161^\circ$ for parahydrophobic PDMS.
All three numbers are valid source values, but only the
first belongs in a static-contact-angle field. Similarly,
the $83$ and $250\,\mathrm{W\,cm^{-2}}$ CHF values discussed
by Chu et al.~\cite{Chu2013CuO} refer to different surfaces;
they do not establish the operating range of one specimen.
These cases motivate checking table headings, variable
definitions, and experimental scope alongside quotations.

For curated releases, we propose domain-expert review of extracted metadata against the source evidence. Reviewers would accept, correct, or reject values, with review decisions and corrections retained alongside source references. Values requiring further clarification would remain flagged for follow-up.

\subsection{Output and evaluation}

Structured records preserve source identifiers,
evidence-location results, original and normalized values,
taxonomy assignments, derived tags, and extractor and
taxonomy version labels. Users can, for example, select
water pool-boiling studies with engineered wettability,
then inspect source evidence and surface conditions before
comparing reported CHF. Paper-level ranges summarize
coverage and do not establish measurement-level pairings.

Evaluation should use independently annotated papers and
report hierarchical tagging precision and recall,
numerical and unit accuracy, semantic evidence support,
unresolved-field rates, and reviewer time. Contact-angle
definitions and surface-specific CHF values provide
concrete evaluation cases. Comparisons among extraction
alone, extraction with automated checks, and extraction
with expert review would quantify each component's
contribution. Independent adjudication of reviewed
assertions would estimate residual error.

\subsection{Cross-study comparisons across six literature categories}
\label{subsec:literature-catalog-comparisons}

\begin{figure*}[p]
\centering
\begin{minipage}[t]{0.482\textwidth}
\centering
\includegraphics[width=\linewidth]{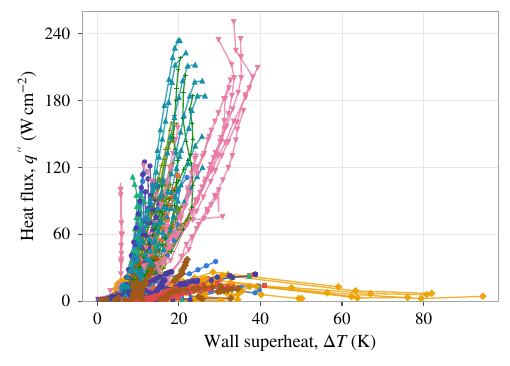}
\caption{\href{https://www.multiphasehub.org/agentic-tool/pool-boiling}{Pool boiling}. The catalog covers 26 papers (1993--2026). The plot shows 127 series from 22 papers.}
\label{fig:lit-pool}
\end{minipage}\hfill
\begin{minipage}[t]{0.482\textwidth}
\centering
\includegraphics[width=\linewidth]{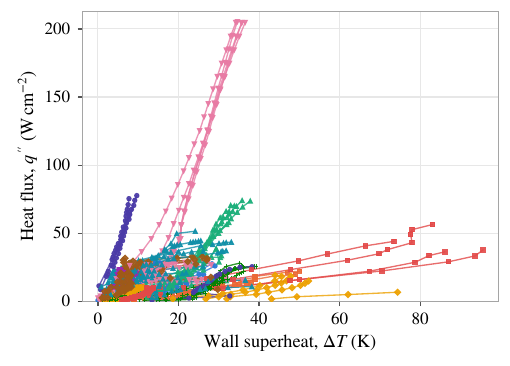}
\caption{\href{https://www.multiphasehub.org/agentic-tool/flow-boiling}{Flow boiling}. The catalog covers 38 papers (2002--2025). The plot shows 156 series from 22 papers.}
\label{fig:lit-flow}
\end{minipage}

\medskip
\begin{minipage}[t]{0.482\textwidth}
\centering
\includegraphics[width=\linewidth]{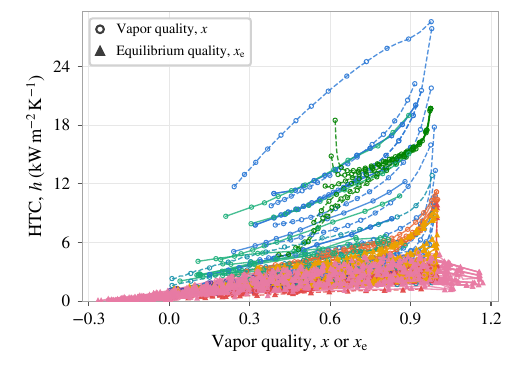}
\caption{\href{https://www.multiphasehub.org/agentic-tool/condensation}{Condensation}. The catalog covers 20 papers (2003--2026). The plot shows 211 series from nine papers.}
\label{fig:lit-condensation}
\end{minipage}\hfill
\begin{minipage}[t]{0.482\textwidth}
\centering
\includegraphics[width=\linewidth]{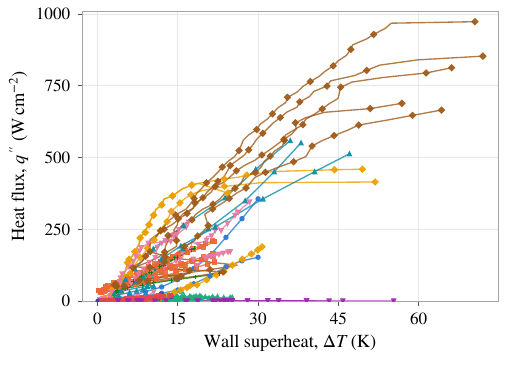}
\caption{\href{https://www.multiphasehub.org/agentic-tool/capillary-boiling}{Capillary boiling}. The catalog covers 19 papers (2002--2024). The plot shows 48 series from 11 papers.}
\label{fig:lit-capillary}
\end{minipage}

\medskip
\begin{minipage}[t]{0.482\textwidth}
\centering
\includegraphics[width=\linewidth]{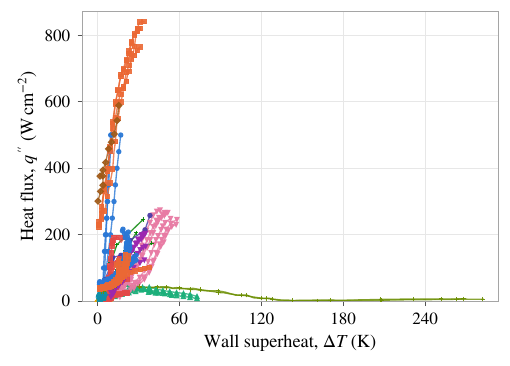}
\caption{\href{https://www.multiphasehub.org/agentic-tool/spray-cooling}{Spray cooling}. The catalog covers 40 papers (1981--2022). The plot shows 58 series from 14 papers.}
\label{fig:lit-spray}
\end{minipage}\hfill
\begin{minipage}[t]{0.482\textwidth}
\centering
\includegraphics[width=\linewidth]{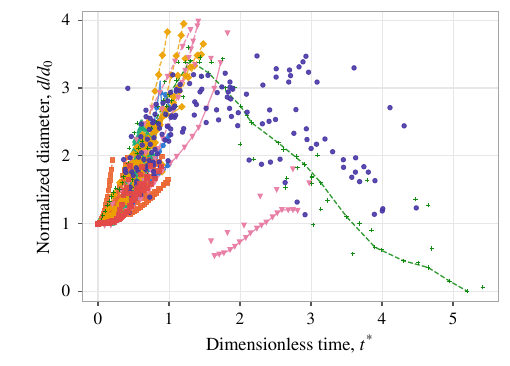}
\caption{\href{https://www.multiphasehub.org/agentic-tool/shock-droplet-interaction}{Shock--droplet interaction}. The catalog covers 11 papers (1969--2025). The displayed dimensionless-time range contains 54 series from eight papers.}
\label{fig:lit-shock}
\end{minipage}
\end{figure*}

The public \href{https://www.multiphasehub.org/agentic-tool}{AgenticTool catalogs} pair searchable study records with digitized series from published figures. Figures~\ref{fig:lit-pool}--\ref{fig:lit-shock} redraw the numerical series displayed by the six category catalogs. The captions give the scope of each catalog and the number of papers that contribute plotted series. The interactive catalogs retain the source paper, figure, and individual series labels.

\paragraph{Pool boiling and flow boiling.}
The pool-boiling catalog spans water, dielectric fluids, refrigerants, and a nanofluid on plain, roughened, nanostructured, porous-coated, and wettability-engineered surfaces, with plotted heat flux up to about 240~W/cm$^2$ and wall superheat up to about 90~K. The flow-boiling catalog covers dielectric fluids, cryogens, and refrigerants, including aerospace and electronics-cooling applications, across conventional, minichannel, and microchannel geometries, with plotted heat flux up to about 200~W/cm$^2$ over a similar range of wall superheat. Both figures show reported heat flux $q''$ as a function of wall superheat $\Delta T$.

\paragraph{Condensation.}
This catalog covers flow condensation of dielectric fluids and refrigerants in circular, horizontal, vertical, and mini- or microchannel geometries, with plotted heat-transfer coefficients up to about 24~kW/(m$^2$K) over vapor qualities from about $-0.3$ to $1.2$. It plots the heat-transfer coefficient $h$ against either vapor quality $x$ or thermodynamic equilibrium quality $x_{\mathrm{e}}$, as defined by the source study. The two quality conventions are marked separately in Fig.~\ref{fig:lit-condensation}.

\paragraph{Capillary boiling and spray cooling.}
The capillary-boiling catalog spans water, dielectric fluids, and cryogens on porous coatings, structured surfaces, and engineered wettability, in both pool and flow configurations, with plotted heat flux up to about 1000~W/cm$^2$ and wall superheat up to about 65~K. The spray-cooling catalog covers water and dielectric sprays, mostly on plain surfaces, for electronics-cooling and aerospace studies, with plotted heat flux up to about 800~W/cm$^2$ and wall superheat up to about 250~K. Their heat-flux plots use the same units and symbols as the pool- and flow-boiling figures.

\paragraph{Shock--droplet interaction.}
This catalog covers water, organic liquids, and a liquid metal in a carrier gas, with applications to propulsion, combustion, and high-speed flight. Figure~\ref{fig:lit-shock} uses the catalog's dimensionless-time view of droplet deformation, plotting normalized diameter $d/d_0$ up to its default display range of $d/d_0\leq 4$ against dimensionless time $t^*$ up to about 5.

For cross-study benchmarking, these plots help identify overlapping ranges and candidate reference datasets. Quantitative comparisons require checking source definitions, operating conditions, and whether a series comes from a measurement, simulation, model, or fit.

\subsection{Digitization check against WebPlotDigitizer}
\label{subsec:compiler_results}

\begin{figure*}[!htpb]
\centering
\includegraphics[width=0.9\textwidth]{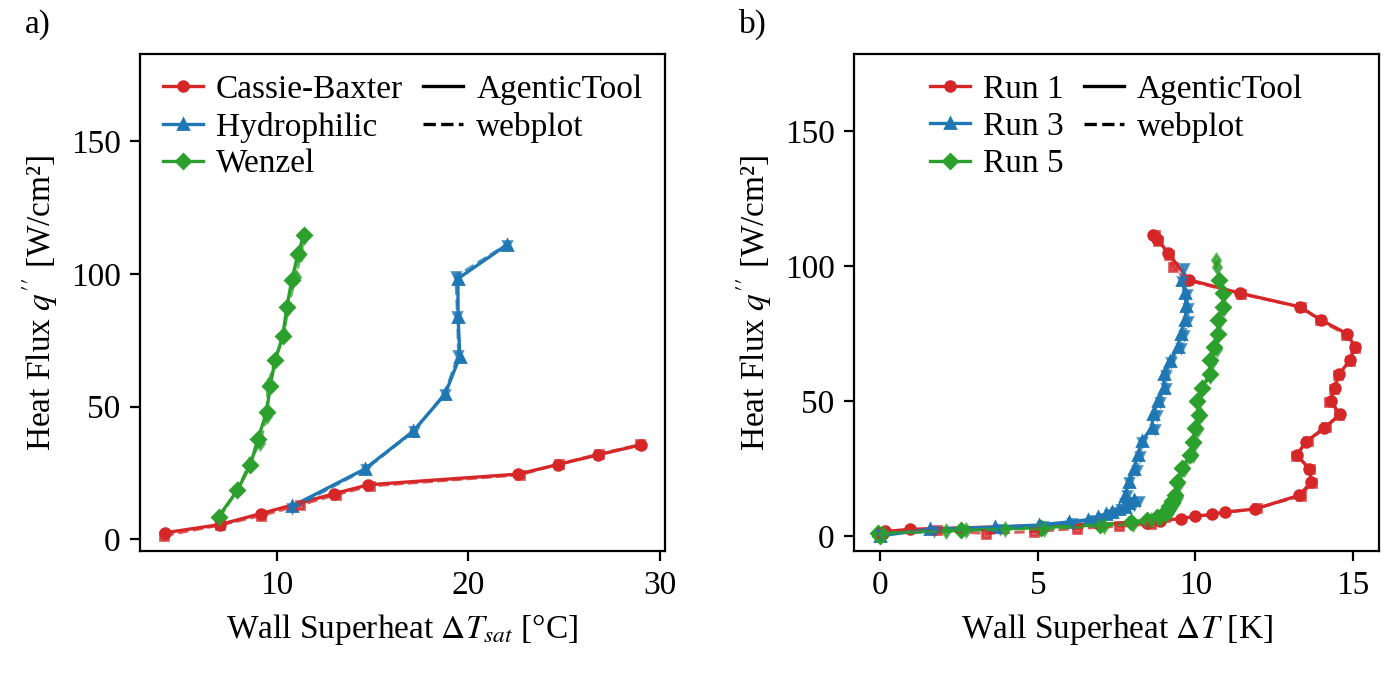}
\caption{Agentic digitizer extraction of boiling curves from two papers, compared with the same curves
digitized manually using WebPlotDigitizer. (a)~Pool boiling on microstructured copper from Fig.~4(a) of Allred
\textit{et al.}~\cite{Allred2018Boiling}:
superhydrophobic surfaces in the Cassie--Baxter and Wenzel states and a hydrophilic bare-copper
reference. (b)~Pool boiling curves from Fig.~5(a) of Berce \textit{et al.}~\cite{berce2024wettability}
: runs~1, 3 and~5 of five
consecutive saturated, atmospheric-pressure runs on a superhydrophilic laser-textured copper surface.
Solid lines show the agentic digitizer extraction and dashed lines the WebPlotDigitizer extraction; colour
identifies the curve within each plot. Wall superheat is shown in degrees Celsius in (a) and kelvin in (b).
Both plots show heat flux $q''$ in $\mathrm{W\,cm^{-2}}$. Conversion to canonical SI heat
flux uses $1\,\mathrm{W\,cm^{-2}}=10^{4}\,\mathrm{W\,m^{-2}}$.}
\label{fig:compiler_boiling-webplot}
\end{figure*}


Figure~\ref{fig:compiler_boiling-webplot} compares curves extracted by the agentic digitizer with manual
digitizations of the same figures in WebPlotDigitizer~\cite{rohatgiwebplotdigitizer}, for Allred \textit{et al.}~\cite{Allred2018Boiling}
(Fig.~4(a)) and Berce \textit{et al.}~\cite{berce2024wettability} (Fig.~5(a)). Both source figures are raster
images, and the tool traced each curve against the printed axes. Comparing against WebPlotDigitizer provides a
human-assisted reference baseline to verify accurate pixel-to-axis calibration on real plots. It also shows whether
automated extraction matches manual digitization quality while removing repetitive manual effort. Since the two methods
sample different numbers of points, we measure agreement as the distance from each agentic digitizer extracted point
to the WebPlotDigitizer curve, normalized by the axis span of the source plot.

The deviation of the $i$-th agentic digitizer extracted point $(x_i,y_i)$ from the WebPlotDigitizer curve is the shortest
distance to that curve in axis-normalized coordinates,
\begin{equation}
  d_i \;=\; \min_{j}\;\min_{t\in[0,1]}
  \sqrt{\left(\frac{x_i - x_j(t)}{L_x}\right)^{2} + \left(\frac{y_i - y_j(t)}{L_y}\right)^{2}},
  \label{eq:digitization_distance}
\end{equation}
where $(x_j(t),\,y_j(t))$ is the straight segment between consecutive WebPlotDigitizer points,
\begin{equation}
  \begin{pmatrix} x_j(t) \\ y_j(t) \end{pmatrix}
  \;=\; (1-t)\begin{pmatrix} x^{\mathrm{W}}_{j} \\ y^{\mathrm{W}}_{j} \end{pmatrix}
  \;+\; t\begin{pmatrix} x^{\mathrm{W}}_{j+1} \\ y^{\mathrm{W}}_{j+1} \end{pmatrix},
  \label{eq:wpd_segment}
\end{equation}
and $L_x$ and $L_y$ are the axis spans of the source plot ($\Delta T$ and $q''$, respectively). The
reported mean deviation over $N$ agentic digitizer extracted points is
$\bar d = (100\%/N)\sum_{i=1}^{N} d_i$.

By this measure the two extractions are close. For Allred \textit{et al.}\
(Fig.~4a, axis span $\approx 26\,\mathrm{K} \times 180\,\mathrm{W\,cm^{-2}}$),
the mean deviation is $0.29\%$ of the axis span and the maximum is $0.65\%$.
In absolute terms, the mean offsets are $0.05\,\mathrm{K}$ in $\Delta T_{sat}$
and $0.3\,\mathrm{W\,cm^{-2}}$ in $q''$, and the maxima are $0.16\,\mathrm{K}$
and $1.1\,\mathrm{W\,cm^{-2}}$. For Berce \textit{et al.}\ (Fig.~5, axis span
$\approx 15\,\mathrm{K} \times 120\,\mathrm{W\,cm^{-2}}$), the mean deviation
is $0.4\%$ and the maximum is $2\%$ of the axis span; the mean offsets are
$0.04\,\mathrm{K}$ and $0.3\,\mathrm{W\,cm^{-2}}$, and the maxima are
$0.16\,\mathrm{K}$ and $2.4\,\mathrm{W\,cm^{-2}}$. The largest Berce
deviations occur along the steep backbending part of the run~1 curve, where a
small horizontal shift corresponds to a large distance in the normalised plane.
Deviations are the nearest-point distance between the two digitized curves,
evaluated in both directions and normalised by the axis span.

\subsection{Dataset discovery and comparison with the MultiphaseHUB Agent}
\label{subsec:dataset-comparison}

The public \href{https://www.multiphasehub.org/assistant}{MultiphaseHUB Agent} takes natural-language questions over dataset records. It complements the \href{https://www.multiphasehub.org/agentic-tool}{AgenticTool} literature catalog, which organizes evidence extracted from publications. In the \href{https://www.multiphasehub.org/videos/agent/agent-demo.mp4}{homepage recording}, the Agent finds datasets tagged for microgravity, displays a side-by-side comparison, plots reported pressure ranges in the boiling collection, and charts catalog counts by pressure category. A dataset can carry multiple category labels, so interpreting the count chart requires checking the category definitions and coverage.

\paragraph{Recorded comparison.}
The demonstration asks the Agent to compare the conditions of BubbleID-Flow and Subatmospheric Boiling. Table~\ref{tab:agent-comparison} transcribes selected fields from the displayed answer. Both records carry a critical heat flux (CHF) coverage label, while their boiling regimes, geometries, and reported pressure information differ. In particular, the answer displays a 10--100~kPa range for Subatmospheric Boiling and leaves the corresponding BubbleID-Flow field blank. The \href{https://zenodo.org/records/22235802}{BubbleID-Flow archive} describes subcooled FC-72 flow boiling in a rectangular passage, and the \href{https://www.multiphasehub.org/datasets/Subatmospheric_Boiling_Dataset}{Subatmospheric Boiling record} describes transient saturated pool boiling. The \href{https://uc-datahub-jyoti.s3.us-east-2.amazonaws.com/Subatmospheric_Boiling_Dataset/README.pdf}{dataset README} documents the latter pressure range. The shared CHF label can help identify candidate datasets. Quantitative comparison requires checking the individual test conditions and how CHF is defined or labeled at the source.

\begin{table}[t]
\centering\small
\caption{Selected fields from the public Agent demonstration. A dash denotes a field left blank in the displayed answer.}
\label{tab:agent-comparison}
\begin{tabularx}{\columnwidth}{@{}>{\raggedright\arraybackslash}p{0.23\columnwidth}YY@{}}
\toprule
\textbf{Field} & \textbf{BubbleID-Flow} & \textbf{Subatmospheric Boiling}\\
\midrule
Bulk liquid & Subcooled & Saturated\\
CHF coverage & Up to CHF & Up to CHF\\
Operation & \textemdash & Transient\\
Geometry & Minichannel & Flat surface, microchannel\\
Stated pressure & \textemdash & 10--100~kPa\\
Conditions source & Zenodo data record & Dataset README\\
\bottomrule
\end{tabularx}
\end{table}

\paragraph{Evaluation.}
The recording documents the user-facing workflow. Systematic assessment would compare Agent responses with independent expert judgments on field and unit accuracy, source support, and recognition of missing or incompatible conditions. Manual catalog search provides a baseline for comparing selection quality and time.

\section{Scientific Compression of Multiphase Transport Data}
\label{sec:compression}

\begin{figure*}[!htpb]
\centering
\includegraphics[width=0.8\textwidth]{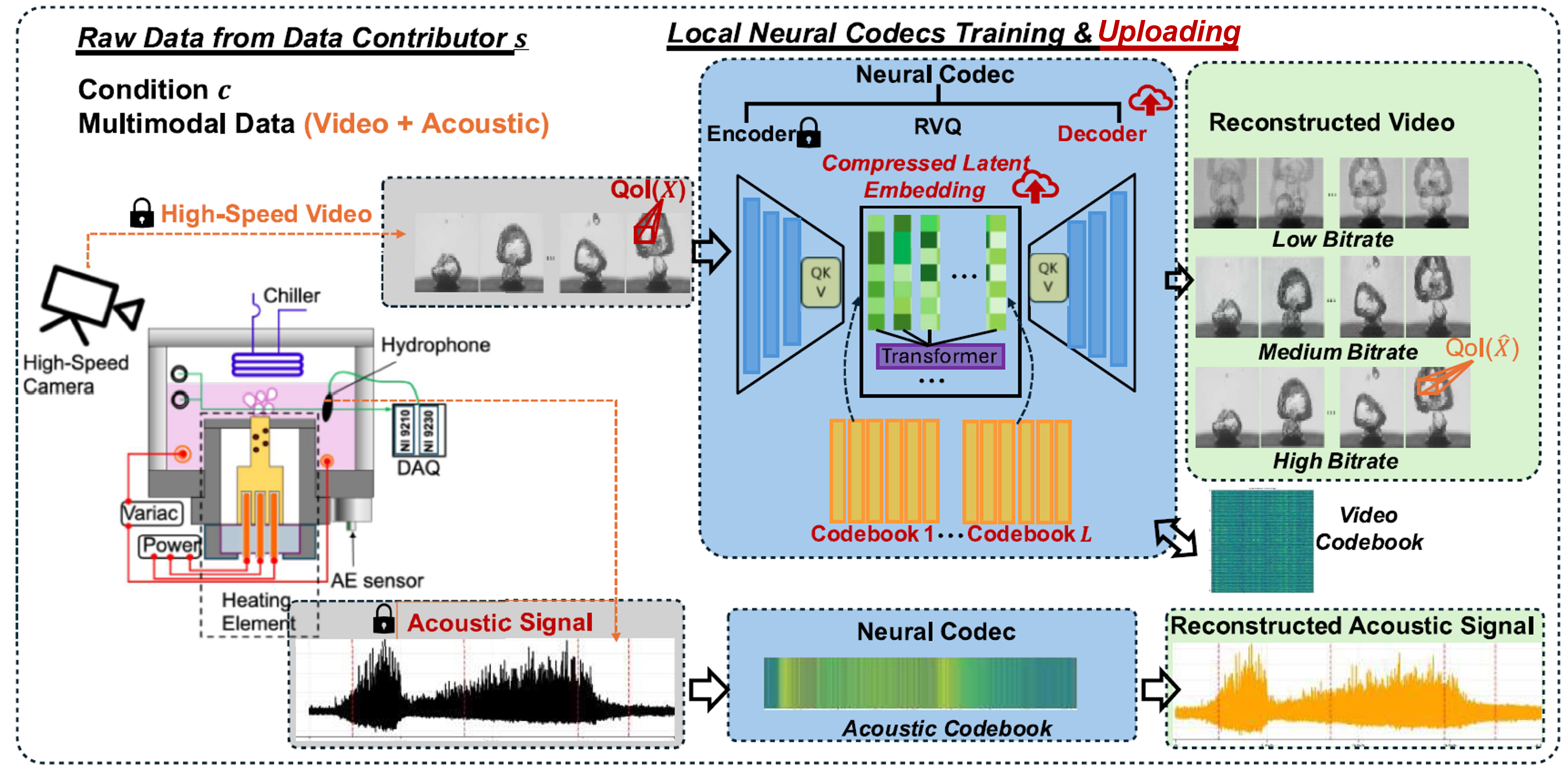}
\caption{Proposed scientific compression workflow for MultiphaseHUB. Modality-specific neural codecs encode measurements into discrete RVQ indices and reconstruct signals using the corresponding codebooks and decoders. Video and acoustic branches are illustrated; the preliminary experiments also include temperature measurements. The bitrate examples and quantities-of-interest annotations illustrate planned rate-utility evaluation. Lock and upload symbols indicate the intended local-processing and sharing arrangement, whose privacy properties require separate evaluation.}
\label{fig:compression_framework}
\end{figure*}

MultiphaseHUB's compression framework uses modality-specific residual vector-quantized variational autoencoders (RVQ-VAEs) to represent temperature signals, hydrophone audio, and boiling videos as discrete indices. Figure~\ref{fig:compression_framework} illustrates the proposed workflow using video and acoustic measurements as examples. Building on vector-quantized representation learning~\cite{vqvae} and residual quantization~\cite{rqvae}, the preliminary implementation evaluates reconstruction from two-stage discrete representations. The video encoder and decoder follow the spatiotemporal VQ-VAE architecture used in VideoGPT~\cite{videogpt}. This study examines reconstruction fidelity, codebook utilization, and payload compression ratios, providing a basis for subsequent evaluation of task-specific scientific quantities.

\subsection{Modality-specific encoding and residual quantization}

For modality $m\in\{\mathrm{temp},\mathrm{audio},\mathrm{video}\}$, an encoder maps a signal segment $x^{(m)}$ to latent vectors $h_j^{(m)}\in\mathbb{R}^{d_m}$ at positions $j=1,\ldots,N_m$. Each position is approximated using two successive quantization stages. With codebook $\mathcal C_{m,\ell}=\{e_{m,\ell,k}\}_{k=1}^{K_m}$ at stage $\ell$, the indices and residuals are
\begin{equation}
\begin{aligned}
r_{j,0}^{(m)} &= h_j^{(m)},\\
z_{j,\ell}^{(m)} &= \underset{k\in\{1,\ldots,K_m\}}{\arg\min}
\big\|r_{j,\ell-1}^{(m)}-e_{m,\ell,k}\big\|_2^2,\\
r_{j,\ell}^{(m)} &= r_{j,\ell-1}^{(m)}-e_{m,\ell,z_{j,\ell}^{(m)}},
\quad \ell=1,2.
\end{aligned}
\label{eq:rvq_indices}
\end{equation}
The decoder reconstructs the input from the summed embeddings,
\begin{equation}
\hat h_j^{(m)}=\sum_{\ell=1}^{2}e_{m,\ell,z_{j,\ell}^{(m)}},
\qquad
\hat x^{(m)}=D_{\phi_m}(\hat h^{(m)}).
\label{eq:rvq_decode}
\end{equation}

RVQ provides a common discrete representation for modality-specific
encoders and decoders. The first stage approximates the encoder output,
and subsequent stages encode the remaining residual. This structure
increases representation capacity through several limited-size codebooks.
Storage requires only the selected indices, together with access to
the corresponding codebooks and decoder.

\paragraph{Stage count and bitrate control.}
For an extended codec with $L_m$ residual stages, reconstruction can
use the first $a$ stages:
\begin{equation}
\begin{aligned}
\hat h_{j,a}^{(m)}
&=
\sum_{\ell=1}^{a}e_{m,\ell,z_{j,\ell}^{(m)}},\\
\hat x_a^{(m)}&=D_{\phi_m}(\hat h_a^{(m)}),
\quad 1\leq a\leq L_m.
\end{aligned}
\label{eq:rvq_active_stages}
\end{equation}
With $K_m$ entries per codebook and fixed-length index packing, the
index payload is
$B_m(a)=N_m a\lceil\log_2 K_m\rceil$ bits per segment,
excluding model storage and side information. Increasing the active
stage count therefore increases the payload and provides additional
residual corrections. The resulting improvement in signal fidelity
and scientific quantities must be evaluated at each setting.

To support decoding at different stage counts, we propose training
with randomly selected active stage counts, following the quantizer-dropout
approach demonstrated in neural audio compression~\cite{soundstream}.
This exposes the decoder to both partial and complete representations.

\paragraph{Incorporating quantities of interest.}
To prioritize information needed for scientific analysis, we propose
augmenting the reconstruction objective with a QoI penalty:
\begin{equation}
\begin{aligned}
\mathcal L_m
&= D_{x,m}(x^{(m)},\hat x_a^{(m)})
 + \lambda_Q D_{Q,m}
 + \beta\mathcal L_{\mathrm{commit},m},\\
D_{Q,m}
&= \sum_k w_{mk}
\frac{
\left\|
g_{mk}(\hat x_a^{(m)},c)
-
g_{mk}(x^{(m)},c)
\right\|_2^2
}{s_{mk}^{2}}.
\end{aligned}
\label{eq:proposed_qoi_loss}
\end{equation}
Here, $D_{x,m}$ measures signal reconstruction error,
$\mathcal L_{\mathrm{commit},m}$ encourages encoder outputs to remain
close to their quantized representations, and $g_{mk}$ computes a
selected scientific quantity using physical context $c$.
The nonnegative weights $w_{mk}$ express task priorities, while
positive scales $s_{mk}$ normalize quantities with different units
or magnitudes. These scales are fixed using physical tolerances
or training data.

Candidate QoIs include projected bubble area, acoustic band energy,
and temperature statistics. Differentiable QoI operators can
contribute directly to training; other quantities can be evaluated
after reconstruction or approximated by a stated differentiable
surrogate. The QoI penalty encourages preservation of selected
quantities, with validation determining the achievable errors
at each active stage count.

The preliminary implementation uses two active stages and separate
codecs for temperature, acoustic signals, and video from one
multimodal boiling dataset. The reported results assess reconstruction,
codebook utilization, and payload compression. Decoding with variable
active stage counts and QoI-guided training are proposed extensions.

\subsection{Boiling dataset and experimental setup}
\label{subsec:compression_setup}
The preliminary compression evaluation uses Boiling-84, a transient run on a copper foam surface from the saturated pool-boiling experiments described by Pandey et al.~\cite{pandey2024multimodal}. The \href{https://doi.org/10.5061/dryad.ksn02v7h2}{Boiling Hysteresis data record} identifies the run and its surface. BoilingBench-1 (Figs.~\ref{fig:bb-raw} and~\ref{fig:bb-processed}) illustrates a separate, subcooled experiment. The saturated experiments used deionized water and five surface configurations: polished copper, a copper microchannel surface, and three electrodeposited copper foams. The experiments included steady-state power steps and transient ramps through critical heat flux. Four embedded thermocouples, a submerged hydrophone, and a high-speed camera recorded the modalities used for compression.

The synchronized temperature, hydrophone, and video files from Boiling-84 yield 10,311 windowed samples using a
nominal window duration of 0.1~s and a stride of 0.05~s. Temperature and
hydrophone samples are separated by 0.000332~s, yielding 324 audio points per window after padding to the nearest multiple of four. Video inputs contain
16 frames (padded from 15) at $128\times128$ spatial resolution and 150 frames per second.
The preliminary experiments use an 80/20 training/validation partition. Because neighboring windows overlap, these results are treated as preliminary within-dataset evidence; source-disjoint generalization requires a partition that prevents shared raw samples across subsets.

\subsection{Preliminary reconstruction results}
\begin{figure*}[!htpb]
\centering
\begin{minipage}[t]{0.48\textwidth}
\centering
\includegraphics[width=\linewidth]{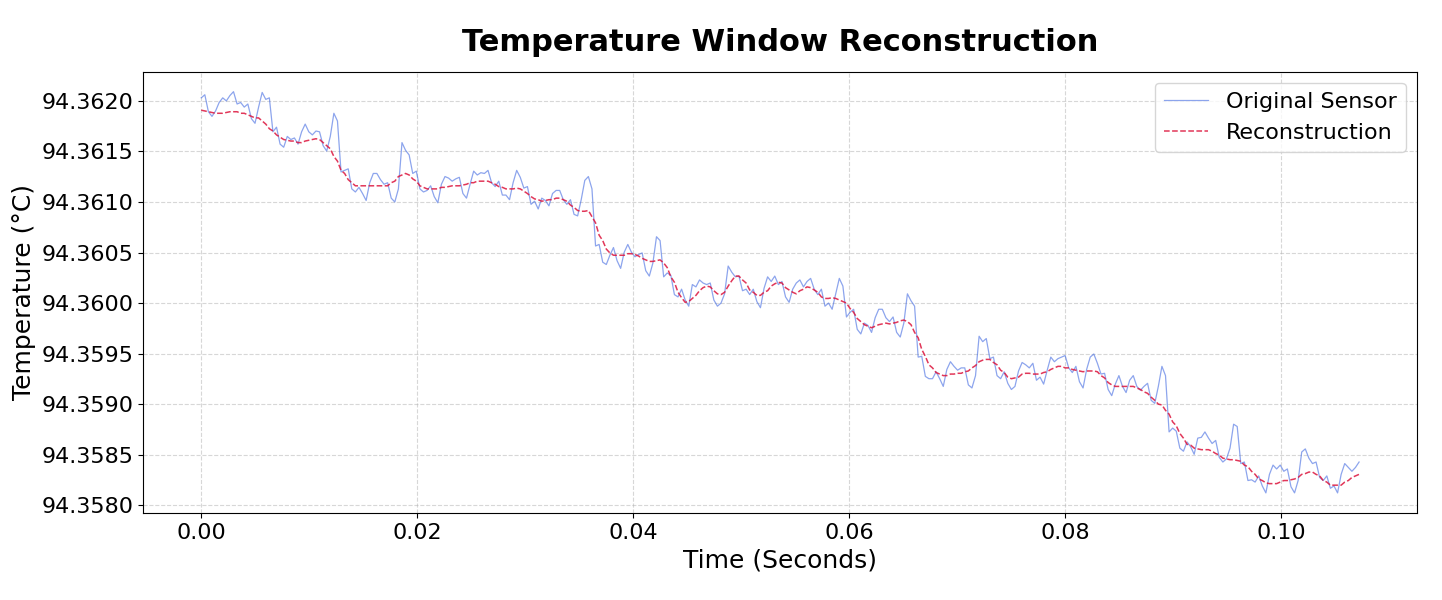}\\[0.5ex]
{\small (a) Temperature}
\end{minipage}\hfill
\begin{minipage}[t]{0.48\textwidth}
\centering
\includegraphics[width=\linewidth]{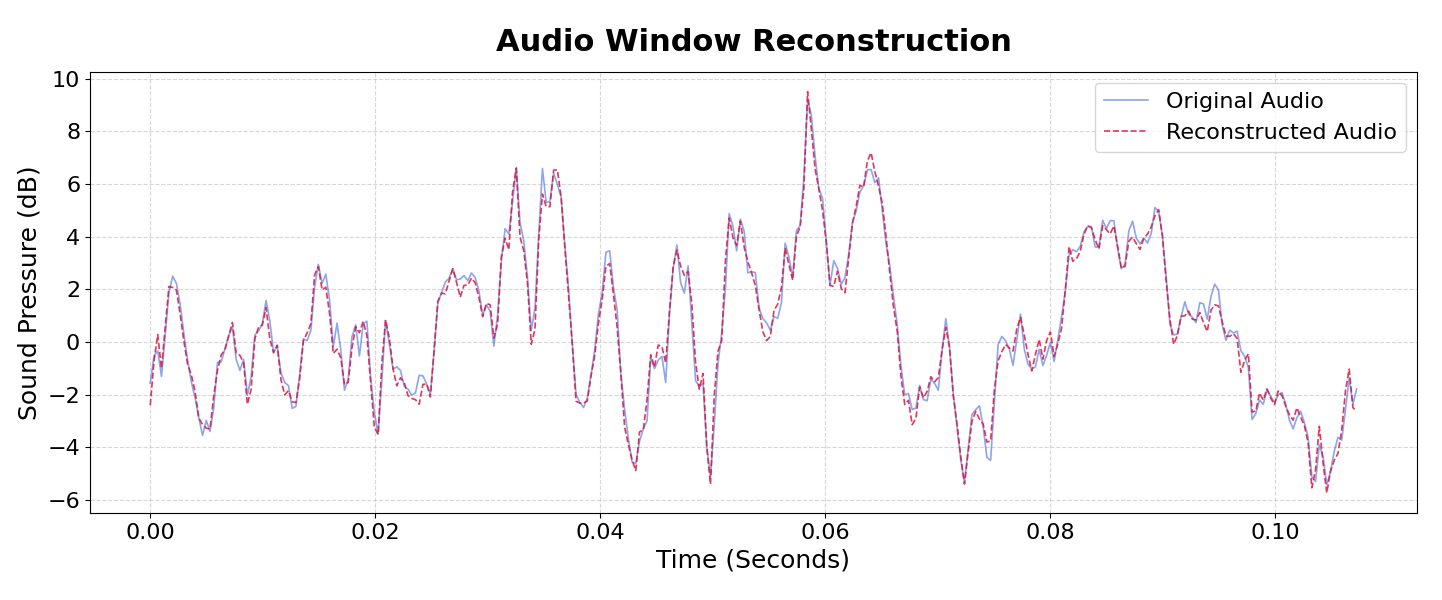}\\[0.5ex]
{\small (b) Hydrophone audio}
\end{minipage}
\par\medskip
\includegraphics[width=\textwidth]{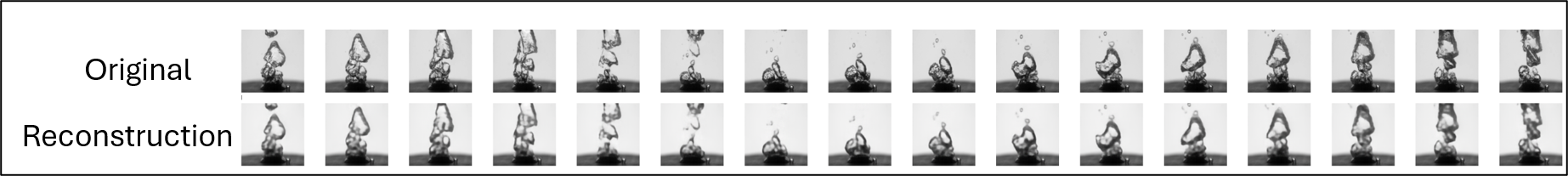}\\[0.5ex]
{\small (c) Boiling video}
\caption{Original and reconstructed evaluation examples for temperature, hydrophone audio, and boiling video using two-stage RVQ-VAEs. The video panel shows original frames above reconstructions, with a sample MSE of 0.0765 at epoch 22.}
\label{fig:rvq_reconstruction}
\end{figure*}

Figure~\ref{fig:rvq_reconstruction} presents original and reconstructed
temperature, acoustic, and video examples from the boiling dataset.
The following results characterize reconstruction quality within this
dataset and provide an initial assessment of the modality-specific codecs.

Table~\ref{tab:rvq_configuration} summarizes the implementation details. Temperature and audio use 64 codes per stage and 48-dimensional latent vectors. The audio encoder uses strided one-dimensional convolutions and a transposed-convolution decoder. The temperature model incorporates input-dependent mean and standard-deviation normalization and replacement of low-usage code vectors. Its replacement threshold is 2\% of the average code usage. Both signal models use a commitment coefficient of 0.25, with exponential-moving-average decay factors of 0.95 for temperature and 0.99 for audio. The video model, derived from VideoGPT, uses three-dimensional convolutions and axial attention. VideoGPT also includes an autoregressive prior in its original formulation~\cite{videogpt}; the results reported here concern the encoder--quantizer--decoder component.

\begin{table}[!htpb]
\centering\small
\caption{RVQ-VAE configurations in the preliminary experiment. Codebook size is per stage; all models use two quantization stages.}
\label{tab:rvq_configuration}
\begin{tabularx}{\columnwidth}{@{}>{\raggedright\arraybackslash}Xccc@{}}
\toprule
 & Temperature & Audio & Video\\
\midrule
Sampling rate & 3,012 Hz & 3,012 Hz & 150 FPS\\
Codes per stage $K_m$ & 64 & 64 & 2,048\\
Latent dimension $d_m$ & 48 & 48 & 256\\
Quantization stages & 2 & 2 & 2\\
Batch size & 64 & 64 & 8\\
\bottomrule
\end{tabularx}
\end{table}

For temperature, the reconstructed trace follows the local trend and much of the small-amplitude variation in the displayed window, with visible smoothing of some rapid fluctuations. For hydrophone audio, the reconstructed waveform reproduces the main positive and negative excursions and their approximate timing in the illustrated segment, while local amplitude and shape differences remain.

\begin{figure*}[!htpb]
\centering
\begin{minipage}[t]{0.48\textwidth}
\centering
\includegraphics[width=\linewidth]{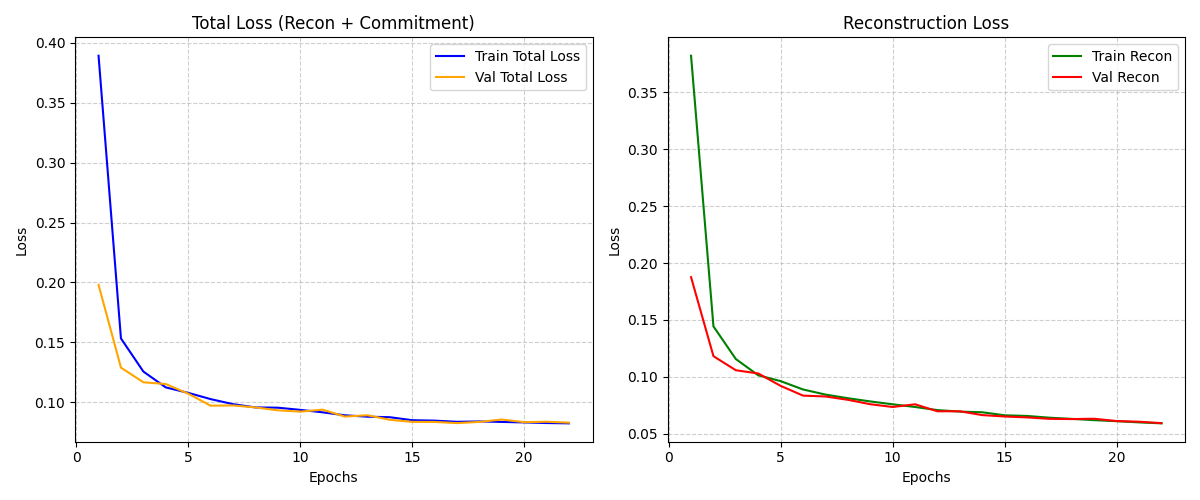}\\[0.5ex]
{\small (a) Video training and validation losses}
\end{minipage}\hfill
\begin{minipage}[t]{0.48\textwidth}
\centering
\includegraphics[width=\linewidth]{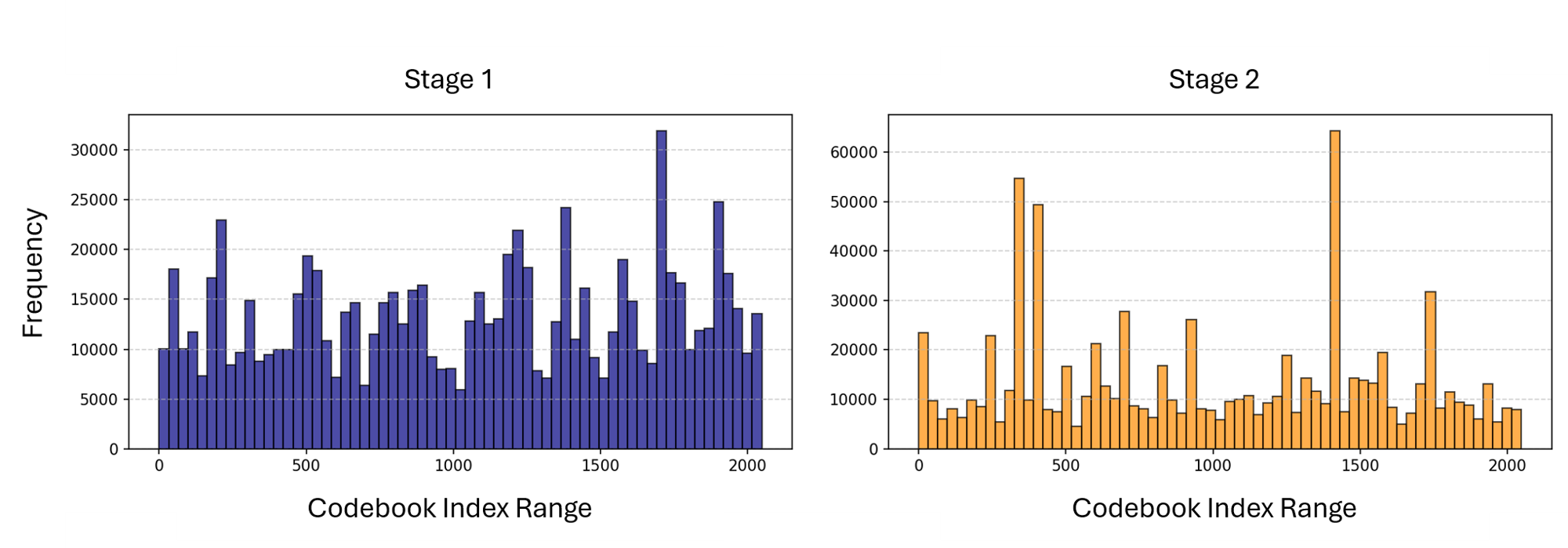}\\[0.5ex]
{\small (b) Video codebook usage}
\end{minipage}
\par\medskip
\begin{minipage}[t]{0.48\textwidth}
\centering
\includegraphics[width=\linewidth]{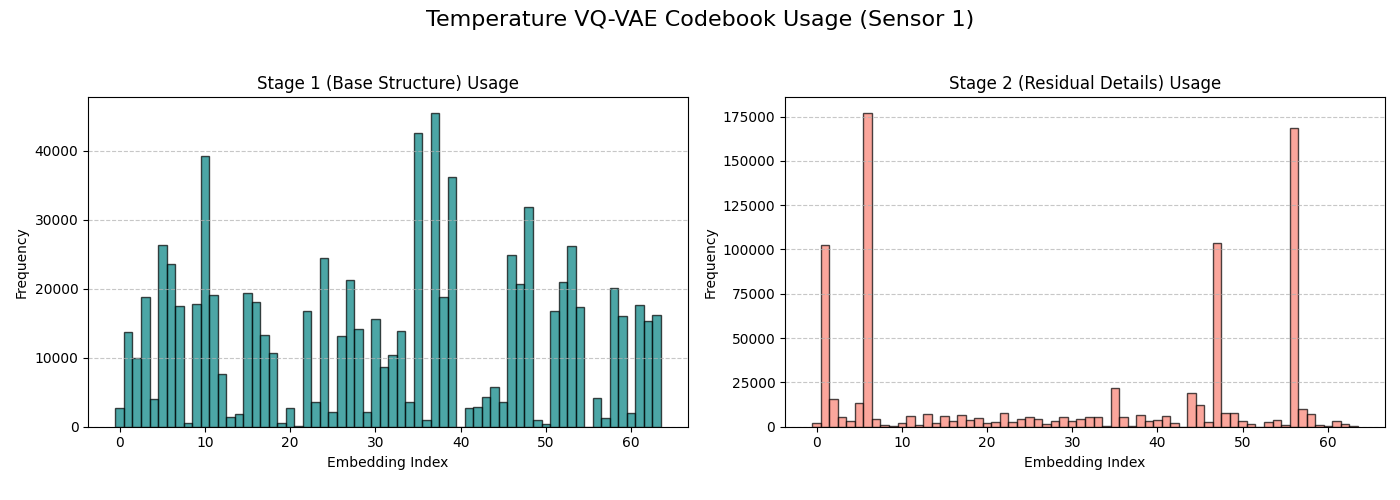}\\[0.5ex]
{\small (c) Temperature codebook usage}
\end{minipage}\hfill
\begin{minipage}[t]{0.48\textwidth}
\centering
\includegraphics[width=\linewidth]{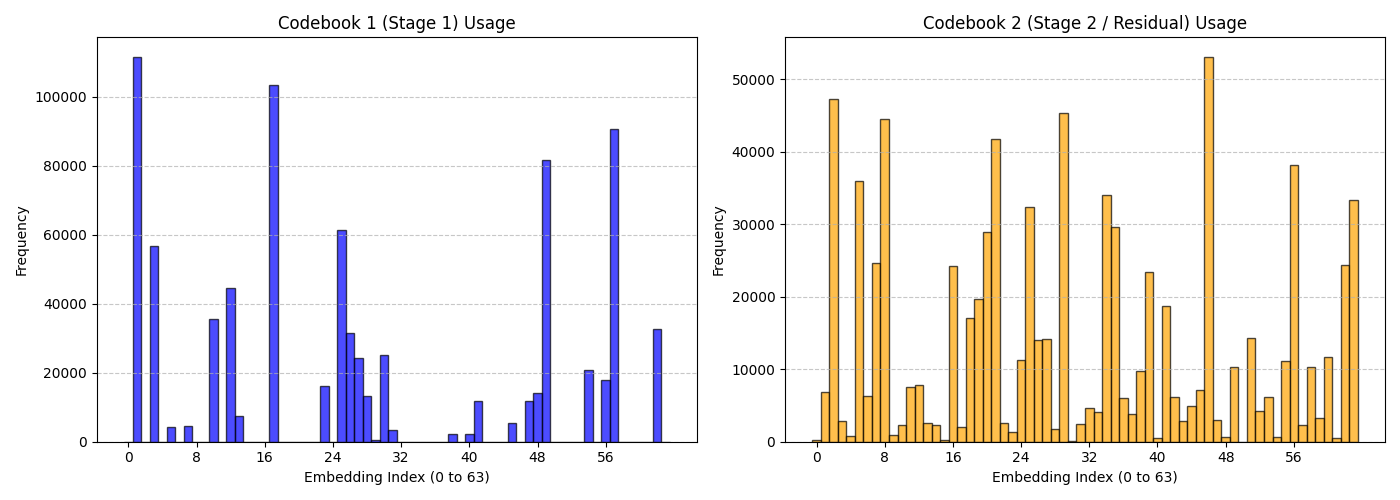}\\[0.5ex]
{\small (d) Hydrophone audio codebook usage}
\end{minipage}
\caption{Training and codebook diagnostics for the two-stage RVQ-VAEs. (a)~Video total loss (reconstruction plus commitment) and reconstruction loss over 22 epochs. (b)~Video codebook usage at epoch 22: 1,739 of 2,048 codes are active in stage 1 and all 2,048 in stage 2; bars aggregate code frequencies over index ranges. (c)~Temperature and (d)~hydrophone audio codebook usage for each stage (64 codes per stage). Active-code fraction is distinct from entropy or effective codebook size.}
\label{fig:rvq_diagnostics}
\end{figure*}

For video, the displayed sequence reconstructs the main bubble structures and their evolution across frames, while fine interface texture and some small features are visibly smoothed. The reconstruction MSE for the illustrated example at epoch 22 is 0.0765, computed on the normalized pixel intensities used as model input. This sample-level metric does not represent a dataset-wide test average.
The video training and validation curves both decrease over the 22 plotted epochs (Figure~\ref{fig:rvq_diagnostics}a).
The observed interface smoothing motivates a subsequent assessment of bubble geometry and event timing.

Codebook utilization provides an additional diagnostic. At video epoch 22, the reported active-code counts are 1,739 of 2,048 in stage 1 (84.9\%) and 2,048 of 2,048 in stage 2 (100\%), as shown in Figure~\ref{fig:rvq_diagnostics}b. These counts show that the representation uses many available codes, although their frequencies remain nonuniform. Temperature stage-2 usage is concentrated on a few codes, and audio exhibits different usage patterns between stages (Figure~\ref{fig:rvq_diagnostics}c,d). Such distributions characterize the learned representations but do not quantify the accuracy gain from the second stage; that gain requires a one-stage versus two-stage reconstruction comparison.

\subsection{Compression ratios and storage accounting}
\label{subsec:compression_ratios}

For modality $m$, the two-stage RVQ representation contains $T_m=2N_m$ code indices across its $N_m$ latent positions. With $K_m$ entries per codebook, each index requires $b_m=\lceil\log_2 K_m\rceil$ bits under fixed-width packing. The index payload and corresponding compression ratio are
\begin{equation}
\begin{aligned}
B_{\mathrm{idx},m}^{\mathrm{bits}} &= T_m b_m,\\
\mathrm{CR}_{\mathrm{payload},m} &=
\frac{B_{\mathrm{raw},m}^{\mathrm{bits}}}{B_{\mathrm{idx},m}^{\mathrm{bits}}}.
\end{aligned}
\label{eq:rvq_payload}
\end{equation}

Table~\ref{tab:compression_ratios} shows that the two-stage RVQ
representations reduce the index payload by more than an order of
magnitude for temperature and audio and by more than two orders of
magnitude for video. Under fixed-width index packing, the corresponding
ratios are $10.7{:}1$ and $279.3{:}1$. The raw reference consists of
324 signal values stored at 32 bits each and 16 video frames of
$128\times128$ pixels with three 8-bit channels per pixel.

These reductions arise from encoding downsampled latent representations
as discrete codebook indices. Each temperature or audio window requires
162 six-bit indices, while each video clip requires 2,048 eleven-bit
indices. Consequently, the video representation occupies only
2,816 bytes per clip, compared with 786,432 bytes for its uncompressed
input.

Across one window of each modality, the total payload decreases from
6,312,192 to 24,472 bits, giving a combined ratio of approximately
$257.9{:}1$. Video dominates the original data volume and therefore
accounts for most of the overall storage saving. With reusable
codebooks and decoders, these compact representations could
substantially reduce repeated transfer of measurement data through
MultiphaseHUB. The reported ratios describe index payloads; deployment
storage also includes the required side information and model files.
Their value for scientific reuse will depend on the quantities
retained at these compression settings.

\begin{table*}[!htpb]
\centering\small
\caption{Per-window compression ratios under fixed-width packing of RVQ indices. Index counts include both quantization stages. Raw values denote scalar signal samples or individual video channel values. Ratios exclude side information, byte-alignment padding, and model storage.}
\label{tab:compression_ratios}
\begin{tabular*}{\textwidth}{@{\extracolsep{\fill}}lrrrrrr@{}}
\toprule
Modality & Raw values & Raw bits & Indices & Bits/index & Index bits & Ratio\\
\midrule
Audio & 324 & 10,368 & 162 & 6 & 972 & $10.7{:}1$\\
Temperature & 324 & 10,368 & 162 & 6 & 972 & $10.7{:}1$\\
Video & 786,432 & 6,291,456 & 2,048 & 11 & 22,528 & $279.3{:}1$\\
\midrule
All three & --- & 6,312,192 & 2,372 & --- & 24,472 & $257.9{:}1$\\
\bottomrule
\end{tabular*}
\end{table*}

\subsection{Integration with MultiphaseHUB}

The compression component produces compact representations of temperature,
acoustic, and video measurements. To support reuse through MultiphaseHUB, each
representation should identify its source dataset, preprocessing,
experimental conditions, and decoder and codebook versions.
These associations allow users to identify the source conditions and
processing steps underlying a reconstructed signal. The preliminary
results provide a starting point for evaluating how compression affects
scientific quantities and for selecting representations suited to a
specified analysis.

\section{Research Roadmap}
\label{sec:roadmap}

\begin{figure*}[!htpb]
\centering
\includegraphics[width=0.9\textwidth]{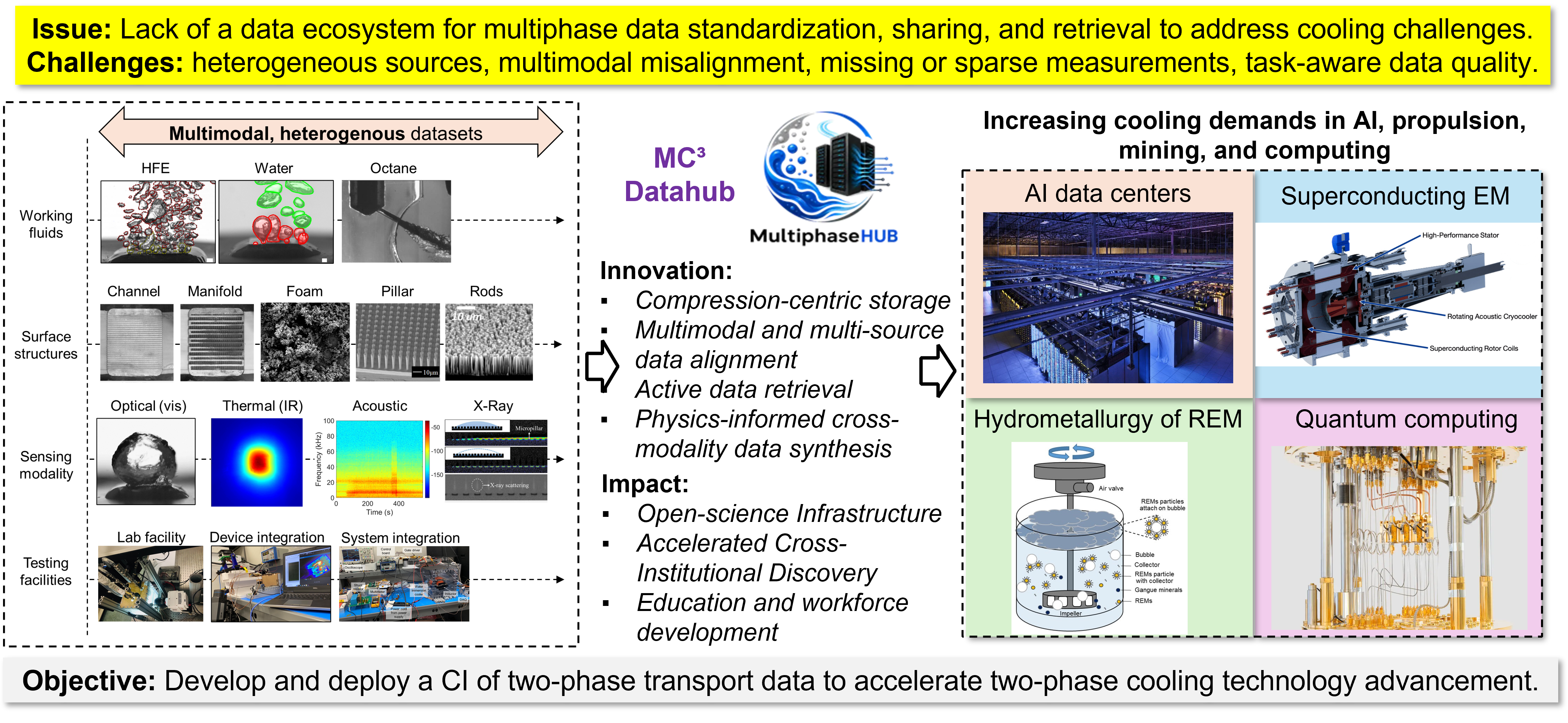}
\caption{Overview of the MultiphaseHUB research roadmap. Heterogenous
multiphase datasets motivate developments in data organization,
multimodal alignment, reconstruction of missing measurements, and
task-aware retrieval and valuation. The application areas illustrate
the broader motivation for the infrastructure and its future
extensions.}
\label{fig:roadmap_overview}
\end{figure*}

The literature-analysis, data extraction and compilation, and compression workflows
support a broader research agenda for multiphase data reuse.
Figure~\ref{fig:roadmap_overview} summarizes this agenda, connecting
heterogeneous data resources with challenges in multimodal integration,
incomplete measurements, and task-dependent data quality. The current
compression study uses one multimodal boiling dataset; extending
evaluation across operating regimes, sensors, and facilities is a
central next step.

\paragraph{Cross-facility harmonization and multimodal alignment.}
Comparisons across studies require consistent interpretation of
physical conditions, variables, units, calibration, and sampling.
Future work will connect literature-derived metadata and compiled
numerical results with experimental measurements to identify
comparable operating regimes. Multimodal alignment will address
differences in time bases, spatial resolution, and sensing mechanisms.
Evaluation across independent studies and facilities will assess
whether these connections support reliable comparison and transfer.

\paragraph{Scientific utility and missing-data reconstruction.}
Task-aware compression will be evaluated through quantities such as
projected bubble area, event timing, acoustic band energy, and
temperature statistics. Applying consistent analysis procedures to
original and reconstructed measurements will establish rate-utility
tradeoffs and guide the choice of active stage count and QoI-based training.
Cross-modal imputation will extend this workflow by using available
measurements and physical context to estimate missing segments or
modalities. The resulting proxies will associate inferred signals
with their uncertainty and supporting observations. Evaluation under
controlled missing-data patterns will assess their usefulness for
downstream analysis.

\paragraph{Task-aware retrieval and dataset valuation.}
The public Agent and the recorded demonstration in Sec.~\ref{subsec:dataset-comparison} establish a starting point for evaluating retrieval and comparison on scientific questions.
Physical metadata, compiled literature data, and extracted features
can support retrieval based on a scientific question. Dataset
valuation will examine the additional information a resource provides
relative to existing evidence, considering operating-regime coverage,
measurement quality, missing information, and reconstruction
uncertainty. These capabilities could help prioritize data curation
and new measurements, identify complementary studies, and reveal
cross-study differences that warrant further physical investigation.

Progress along these directions will require broader seed resources,
reproducible evaluation workflows, and community contributions.
Versioned source links, extraction methods, processing settings,
and access conditions will support verification and reuse as the
infrastructure expands.

\section{Conclusions}

MultiphaseHUB links distributed multiphase transport data to physical context, provenance, and reuse conditions. The public MultiphaseHUB Agent supports dataset discovery, side-by-side comparison, and coverage charts. The recorded boiling comparison illustrates why regime, geometry, and reported pressure matter for reuse. The public literature catalog supports cross-study inspection through metadata extraction and figure digitization.

The public literature catalogs cover six categories, with digitized series shown in Figs.~\ref{fig:lit-pool}--\ref{fig:lit-shock}. Section~\ref{subsec:compiler_results} compares automated digitization with manual WebPlotDigitizer traces of two published pool-boiling figures.

A preliminary study using one multimodal boiling dataset demonstrated
two-stage RVQ representations for temperature, hydrophone audio,
and video. The packed-index compression ratios are approximately
$10.7{:}1$ for temperature and audio and $279.3{:}1$ for video,
excluding model storage and side information. Reconstruction examples
and codebook utilization provide initial evidence for these compact
representations.

These results are preliminary. The compression study uses one
transient run on a copper foam surface, with overlapping windows divided 80/20
into training and validation subsets; reconstruction is characterized
by examples rather than dataset-wide metrics, and the preservation of
QoIs has not yet been evaluated. Metadata extraction has not been
scored against independent annotations, and digitization accuracy was
assessed for two figures only. Addressing these gaps, together with
the roadmap in Section~\ref{sec:roadmap}, is the next step toward
evaluating scientific utility, integrating incomplete multimodal
measurements, and supporting physically meaningful cross-study
discovery.

\section*{Data and Code Availability}
The public platform is available at \url{https://www.multiphasehub.org/}, with entry points for the \href{https://www.multiphasehub.org/datasets}{dataset catalog}, the \href{https://www.multiphasehub.org/assistant}{MultiphaseHUB Agent}, the \href{https://www.multiphasehub.org/agentic-tool}{literature catalog}, and \href{https://www.multiphasehub.org/software}{research software}. The \href{https://www.multiphasehub.org/videos/agent/agent-demo.mp4}{public Agent recording} documents the example in Sec.~\ref{subsec:dataset-comparison}. The catalog links to original repositories and licenses. BoilingBench-1 is part of the \href{https://doi.org/10.5281/zenodo.22131859}{BoilingBench-Multimodal Lite archive}~\cite{hu2026boilingbench}. The saturated pool-boiling dataset described by Pandey et al.~\cite{pandey2024multimodal} has a \href{https://doi.org/10.7910/DVN/6GLGC6}{Harvard Dataverse record}. Transient experiments including Boiling-84 are also indexed in the \href{https://ned3.uark.edu/datasets/}{NED3 laboratory catalog} under Boiling Hysteresis, with a \href{https://doi.org/10.5061/dryad.ksn02v7h2}{Dryad companion archive}. BubbleID source data are available from Dryad~\cite{bubbleiddata}.

The literature extraction and figure digitization code is available at \url{https://github.com/dr-jppanda/AgenticTool-MultiPhaseHUB}~\cite{Panda2026AgenticToolMultiPhaseHUB}.

\bibliographystyle{unsrt}
\bibliography{multiphasehub_references}

\onecolumn
\appendix
\section{Dataset inventory by intended role}
\label{app:dataset-roles}

Table~\ref{tab:dataset-roles} gives the inventory snapshot's resource-level labels by application area. \emph{Train + Eval} marks candidates for both model development and evaluation, such as BoilingBench-Multimodal. \emph{Eval} marks resources primarily intended for evaluation, such as the NREL ESIF cooling-system data. \emph{Context only} includes workload traces that describe operating scenarios, and \emph{Supporting} includes artifacts such as twisted-pair specimen-preparation designs. Section~\ref{subsec:dataset-roles} explains how researchers can use particular versions or subsets as evidence, references, or tests for a specific hypothesis.

\begin{table}[!ht]
\centering\small
\setlength{\tabcolsep}{6pt}
\renewcommand{\arraystretch}{1.30}
\definecolor{inventoryNavy}{HTML}{192B3E}
\definecolor{inventoryStripe}{HTML}{F3F6F9}
\definecolor{inventoryTrain}{HTML}{E1F1E8}
\definecolor{inventoryEval}{HTML}{E3EDF8}
\definecolor{inventoryContext}{HTML}{EEEAF6}
\definecolor{inventoryPending}{HTML}{F7ECD8}
\definecolor{inventoryTotal}{HTML}{DEE7EF}
\newcommand{\inventoryhead}[1]{\textcolor{white}{\bfseries #1}}
\newcommand{\inventorycount}[2]{\ifnum#2>0\relax\cellcolor{#1}\textbf{#2}\else\textcolor{gray}{\textemdash}\fi}
\caption{Dataset inventory by application area and intended role.}
\label{tab:dataset-roles}
\rowcolors{2}{inventoryStripe}{white}
\begin{tabularx}{\textwidth}{Yrrrrrr}
\rowcolor{inventoryNavy}
\rule[-1.30ex]{0pt}{6.2ex}\inventoryhead{Application area} &
\inventoryhead{Records} & \inventoryhead{\shortstack[r]{Train +\\Eval}} &
\inventoryhead{Eval} & \inventoryhead{\shortstack[r]{Context\\only}} &
\inventoryhead{Supporting} & \inventoryhead{Pending}\\
\midrule
Boiling and condensation & 47 & \inventorycount{inventoryTrain}{41} & \inventorycount{inventoryEval}{2} & \inventorycount{inventoryContext}{0} & \inventorycount{inventoryContext}{0} & \inventorycount{inventoryPending}{4}\\
Data center and electronics cooling & 13 & \inventorycount{inventoryTrain}{7} & \inventorycount{inventoryEval}{4} & \inventorycount{inventoryContext}{2} & \inventorycount{inventoryContext}{0} & \inventorycount{inventoryPending}{0}\\
Hypersonics and propulsion & 5 & \inventorycount{inventoryTrain}{5} & \inventorycount{inventoryEval}{0} & \inventorycount{inventoryContext}{0} & \inventorycount{inventoryContext}{0} & \inventorycount{inventoryPending}{0}\\
Spacecraft thermal control and electrification & 4 & \inventorycount{inventoryTrain}{1} & \inventorycount{inventoryEval}{2} & \inventorycount{inventoryContext}{0} & \inventorycount{inventoryContext}{1} & \inventorycount{inventoryPending}{0}\\
Droplet impact and particle deposition & 14 & \inventorycount{inventoryTrain}{12} & \inventorycount{inventoryEval}{2} & \inventorycount{inventoryContext}{0} & \inventorycount{inventoryContext}{0} & \inventorycount{inventoryPending}{0}\\
Industrial multiphase processes & 18 & \inventorycount{inventoryTrain}{16} & \inventorycount{inventoryEval}{2} & \inventorycount{inventoryContext}{0} & \inventorycount{inventoryContext}{0} & \inventorycount{inventoryPending}{0}\\
Cryogenics and quantum technologies & 5 & \inventorycount{inventoryTrain}{4} & \inventorycount{inventoryEval}{1} & \inventorycount{inventoryContext}{0} & \inventorycount{inventoryContext}{0} & \inventorycount{inventoryPending}{0}\\
Semiconductor devices and packaging & 4 & \inventorycount{inventoryTrain}{2} & \inventorycount{inventoryEval}{2} & \inventorycount{inventoryContext}{0} & \inventorycount{inventoryContext}{0} & \inventorycount{inventoryPending}{0}\\
Geoscience & 15 & \inventorycount{inventoryTrain}{12} & \inventorycount{inventoryEval}{3} & \inventorycount{inventoryContext}{0} & \inventorycount{inventoryContext}{0} & \inventorycount{inventoryPending}{0}\\
Critical materials and mineral recovery & 17 & \inventorycount{inventoryTrain}{4} & \inventorycount{inventoryEval}{13} & \inventorycount{inventoryContext}{0} & \inventorycount{inventoryContext}{0} & \inventorycount{inventoryPending}{0}\\
Manufacturing & 9 & \inventorycount{inventoryTrain}{9} & \inventorycount{inventoryEval}{0} & \inventorycount{inventoryContext}{0} & \inventorycount{inventoryContext}{0} & \inventorycount{inventoryPending}{0}\\
Energy storage & 3 & \inventorycount{inventoryTrain}{3} & \inventorycount{inventoryEval}{0} & \inventorycount{inventoryContext}{0} & \inventorycount{inventoryContext}{0} & \inventorycount{inventoryPending}{0}\\
Biomanufacturing & 2 & \inventorycount{inventoryTrain}{2} & \inventorycount{inventoryEval}{0} & \inventorycount{inventoryContext}{0} & \inventorycount{inventoryContext}{0} & \inventorycount{inventoryPending}{0}\\
\midrule
\rowcolor{inventoryTotal}
\textbf{Total} & \textbf{156} & \textbf{118} & \textbf{31} & \textbf{2} & \textbf{1} & \textbf{4}\\
\bottomrule
\end{tabularx}
\par\smallskip
\begin{minipage}{\textwidth}
\footnotesize
A dash denotes zero. Pending indicates a role to be confirmed. The two particle-deposition rows share a hub entry and are counted once. The inventory totals include two boiling records marked as unlisted.
\end{minipage}
\end{table}

\end{document}